\documentclass[aps,prd, twocolumn,superscriptaddress,floatfix,preprintnumbers,nofootinbib]{revtex4-1}

\usepackage[utf8]{inputenc}
\usepackage[pdftex,bookmarks,linktocpage,pdfpagelabels,plainpages=false,hyperfigures,linkcolor=blue,citecolor=blue]{hyperref}
\hypersetup{colorlinks=true}
\usepackage{subfig}
\usepackage{bm}
\usepackage{esvect}
\usepackage{verbatim}
\usepackage{multirow}

\usepackage[left]{lineno}
\usepackage{blindtext}
\pdfoutput=1
\usepackage[T1]{fontenc}
\usepackage{graphicx}
\usepackage{epsfig}
\usepackage{amsmath}
\usepackage{amsfonts}
\usepackage{amssymb}
\usepackage{braket}
\usepackage{cancel}
\usepackage{pstricks}
\usepackage{color}
\usepackage{feynmf}
\usepackage[normalem]{ulem}
\usepackage{epstopdf}
\usepackage{soul}
\usepackage{cancel}
\usepackage{multirow}
\usepackage{mathrsfs}

\makeatletter
\renewcommand\@makecaption[2]{%
  \par
  \vskip\abovecaptionskip
  \begingroup
   \small\rmfamily
    \begingroup
     \samepage
     \flushing
     \let\footnote\@footnotemark@gobble
     \@make@capt@title{#1}{#2}\par
    \endgroup
  \endgroup
  \vskip\belowcaptionskip
}
\makeatother

\definecolor{ao(english)}{rgb}{0.0, 0.5, 0.0}

\usepackage{braket}
\usepackage{cancel}
\usepackage{setspace}
\usepackage{pstricks}
\usepackage{xcolor}
\usepackage{float}
\usepackage{mathtools}
\usepackage{multirow}
\usepackage{booktabs}
\usepackage{enumitem}

\def\gsim{\lower0.5ex\hbox{$\:\buildrel >\over\sim\:$}}
\def\lsim{\lower0.5ex\hbox{$\:\buildrel <\over\sim\:$}}

\begin{document}

\title{Neutral and doubly charged scalars at future lepton colliders}

\author{Fang Xu}
\email{xufang@wustl.edu}
\affiliation{Department of Physics, Washington University, St.~Louis, MO 63130, USA}


\begin{abstract}
Many new physics scenarios beyond the Standard Model (BSM)  necessitate the existence of new neutral and/or charged scalar fields, which might couple to the SM charged leptons (but not hadrons) and, thus, can give rise to BSM signals while evading strong constraints mostly coming from the hadronic sector. I show that future lepton colliders provide a clean environment to probe these leptophilic new scalars via multilepton final states, including some interesting lepton flavor violating channels. I also study the kinematic distributions of the final state leptons to distinguish the BSM contributions from neutral and doubly charged scalars giving rise to the same final state, as well as from the irreducible SM background.

\end{abstract}

\maketitle
\flushbottom


\section{Introduction}
\label{sec:intro}
The Standard Model (SM) of particle physics has been tremendously successful in explaining a wide variety of experimental results and characterizing the fundamental particles and their interactions. However, there are still many unresolved issues that need to extend the SM by introducing new particles and interactions, such as the Higgs sector extension models, which introduce additional scalar fields~\cite{Gunion:1989we,Gunion:2002in,Carena:2002es,Lykken:2010mc}.

The neutral and doubly charged Higgs bosons are hypothetical particles predicted by certain extensions of the SM, such as the Left-Right Symmetric Model (LRSM)~\cite{Pati:1974yy,Mohapatra:1974gc,Senjanovic:1975rk,Mohapatra:1977mj,BhupalDev:2018vpr}, Two Higgs Doublet Model (2HDM)~\cite{Lee:1973iz,Gunion:1989we,Haber:1984rc,Gunion:1984yn,Branco:2011iw,Black:2022wbg} and Higgs Triplet Model (HTM)~\cite{Konetschny:1977bn,Magg:1980ut,Schechter:1980gr,Cheng:1980qt,Lazarides:1980nt}. The new neutral (dubbed as `$H_{3}$') and doubly charged (dubbed as `$H^{\pm\pm}$') scalar fields might couple to the SM charged leptons through Yukawa interactions:
\begin{align}
    \mathcal{L}_{H_{3}} \ \supset \ Y_{\alpha\beta}\overline{\ell_{\alpha}}H_{3}\ell_{\beta}+{\rm H.c.},
\label{eq:Largangian1}
\end{align}
\begin{align}
    \mathcal{L}_{H^{++}} \ \supset \ Y_{\alpha\beta}\overline{\ell^{C}_{\alpha}}H^{++}\ell_{\beta}+{\rm H.c.}.
\label{eq:Largangian2}
\end{align}

For example, in the LRSM, the physical fields $H_{3}$ and $H^{\pm\pm}$ come from the triplet Higgs fields $\Delta_{\rm L,R}$: $H_{3}\equiv {\rm Re}(\Delta^{0})$ and $H_{\rm L,R}^{\pm\pm}\equiv \Delta^{\pm\pm}_{\rm L,R}$, where
\begin{equation}
    \Delta_{\rm L,R} \ = \ \begin{pmatrix}\Delta_{\rm L,R}^{+}/\sqrt{2}&\Delta_{\rm L,R}^{++}\\\Delta_{\rm L,R}^{0}&-\Delta_{\rm L,R}^{+}/\sqrt{2}\end{pmatrix},
\label{eq:triplet Higgs}
\end{equation}
and the triplet $\Delta_{\rm L,R}$ couples to the lepton doublets $L_{{\rm L},\alpha} = (\nu_{\rm L},e_{\rm L})_{\alpha}^{\sf T}$ and $L_{{\rm R},\alpha} = (\nu_{\rm R},e_{\rm R})_{\alpha}^{\sf T}$ through Yukawa interactions
\begin{align}
    \mathcal{L}_{\rm Y} \ \supset \ & Y_{{\rm L},\alpha\beta}L_{{\rm L},\alpha}^{\sf T}C^{-1}\sigma_{2}\Delta_{\rm L}L_{{\rm L},\beta} \nonumber \\
    & +Y_{{\rm R},\alpha\beta}L_{{\rm R},\alpha}^{\sf T}C^{-1}\sigma_{2}\Delta_{\rm R}L_{{\rm R},\beta}+{\rm H.c.},
\label{eq:Lagrangian LRSM}
\end{align}
where $\alpha$ and $\beta$ denote the lepton flavor $e,\mu,{\rm \ or \ }\tau$ and $C$ is the charge conjugation matrix.

These Yukawa interactions are important in addressing the (smallness) of the neutrino masses, the baryon asymmetry of the Universe and are responsible for the origin of the electroweak symmetry breaking. The observation of the Yukawa couplings could have important implications for different beyond the Standard Model (BSM) models, such as theories of neutrino mass and baryogenesis.

If the mass scale of the scalars is less than a few TeV, the lepton colliders (with the $\sqrt{s} \sim$ TeV scale) are especially well suited for detections of the scalar fields as well as the study of the corresponding BSM scenarios because they provide clean backgrounds and signals compared to hadron colliders. Lepton colliders can also be run at a fixed center-of-mass energy, which allows for precise control of the collision energy and provides a spectacular chance for the study of the new particles at high precision, such as their masses, couplings, and decay modes.

In this work, ignoring the actual detailed form of the Yukawa interactions, I simply treat the overall Yukawa couplings as model-independent observables and study their discovery prospect at future lepton colliders such as the International Linear Collider (ILC)~\cite{ILC:2013jhg,Evans:2017rvt} with a center-of-mass energy of 1.0 TeV and the Compact Linear Collider (CLIC)~\cite{CLICPhysicsWorkingGroup:2004qvu,CLIC:2016zwp} with center-of-mass energies of 1.5 TeV and 3 TeV. The final states in the collider searches will be fairly similar as both neutral and doubly charged scalars couple to SM charged leptons. In this paper, I also make a detailed investigation, outlining the differences between their final states under various cases, along with their distinctive dilepton invariant mass distributions.

The rest of the paper is organized as follows: in Section~\ref{sec:theoretical}, I briefly discuss the characteristic cases and the related experimental constraints that are used in this paper. Section~\ref{sec:collider} presents the detailed analysis of the collider signal considered in this paper, as well as the background evaluation and different cuts that are used for different cases. The results and other discussions are summarized in Section~\ref{sec:results} and Section~\ref{sec:discussions}. Section~\ref{sec:conclusions} gives the conclusions. The related invariant mass distributions are shown in the Appendix~\ref{sec:invariant mass}.

\section{Theoretical analysis}
\label{sec:theoretical}
At lepton colliders, the interaction terms presented in Eq.~\eqref{eq:Largangian1} and Eq.~\eqref{eq:Largangian2} will produce highly comparable final states consisting of multiple charged leptons. To demonstrate the differences and for simplicity, I only consider the $e, \mu$ sector of the Yukawa matrices and further assume the diagonal and off-diagonal couplings are equal separately: $|Y_{e e}| = |Y_{\mu \mu}|$ and $|Y_{e \mu}| = |Y_{\mu e}|$.

Because of the extremely strong constraints of $\mu \to e e \Bar{e} \ (< 1.0 \times 10^{-12} {\rm \ at \ 90\% \ CL})$~\cite{SINDRUM:1987nra} and $\mu \to e \gamma \ (< 4.2 \times 10^{-13} {\rm \ at \ 90\% \ CL})$~\cite{MEG:2016leq}, the diagonal and off-diagonal terms cannot be large at the same time. Otherwise, these lepton flavor violating (LFV) rare lepton decay processes can happen at the tree level (for $\mu \to e e \Bar{e}$) and one-loop level (for $\mu \to e \gamma$) with a neutral or doubly charged mediator in the diagrams.

Thus, to describe the differences between neutral and doubly charged scalars and the differences between diagonal and off-diagonal Yukawa couplings in each scalar interaction term, I consider four characteristic cases in this work:
\begin{enumerate}
\item [(a)] Neutral scalar $H_{3}$ with nonzero diagonal Yukawa couplings $|Y_{e e}| = |Y_{\mu \mu}|$.
\item [(b)] Neutral scalar $H_{3}$ with nonzero off-diagonal Yukawa couplings $|Y_{e \mu}| = |Y_{\mu e}|$.
\item [(c)] Doubly charged scalar $H^{\pm\pm}$ with nonzero diagonal Yukawa couplings $|Y_{e e}| = |Y_{\mu \mu}|$.
\item [(d)] Doubly charged scalar $H^{\pm\pm}$ with nonzero off-diagonal Yukawa couplings $|Y_{e \mu}| = |Y_{\mu e}|$.
\end{enumerate}

From now on, we will use a convention in this paper that the letters (a), (b), (c), and (d) appearing in the equations, figures, and tables below, respectively, correspond to the four cases (a), (b), (c), and (d) presented here.

For the possible experimental constraints in the four cases, I use the data from the rare LFV decays $\ell_\alpha \to \ell_\beta \ell_\gamma \ell_\delta$, $\ell_\alpha \to \ell_\beta \gamma$~\cite{Workman:2022ynf,HFLAV:2016hnz}, the muonium oscillation~\cite{Willmann:1998gd}, the LEP $e^+ e^- \to \ell^+ \ell^-$~\cite{DELPHI:2005wxt}, and the LHC  multilepton~\cite{ATLAS-CONF-2021-011,ATLAS:2022pbd}.

\begin{figure*}[t!]
    \centering
    \setcounter{subfigure}{0}\renewcommand{\thesubfigure}{b}\makeatletter
    \subfloat[${H_3}$, $|Y_{e\mu}| \neq 0$]{\includegraphics[width=0.35\textwidth]{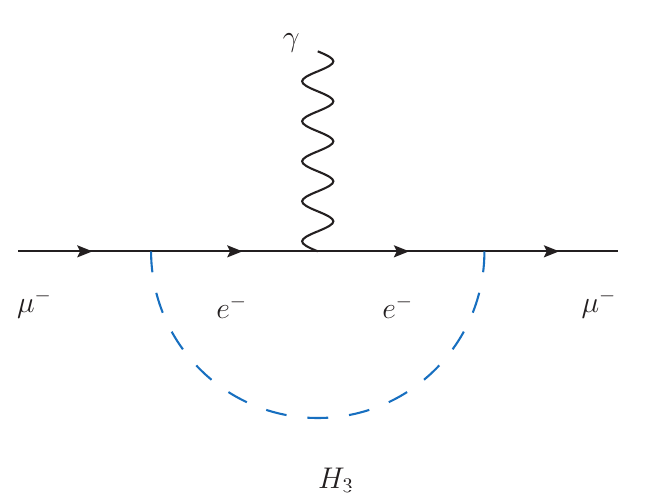}}
    \setcounter{subfigure}{0}\renewcommand{\thesubfigure}{d\arabic{subfigure}}\makeatletter
    \subfloat[${H^{\pm\pm}}$, $|Y_{e\mu}| \neq 0$]{\includegraphics[width=0.35\textwidth]{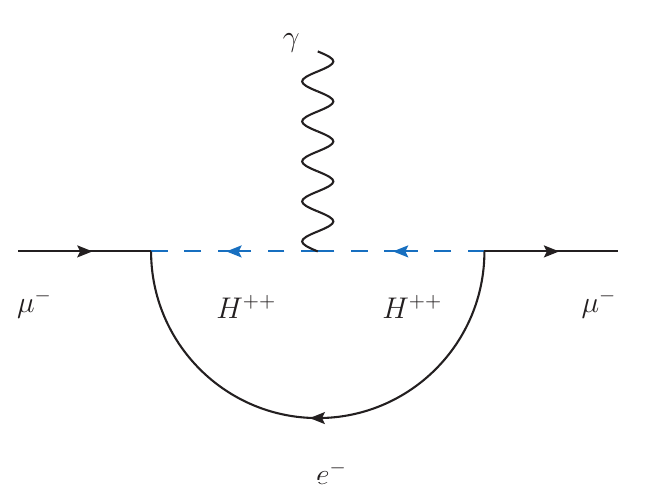}}
    \subfloat[${H^{\pm\pm}}$, $|Y_{e\mu}| \neq 0$]{\includegraphics[width=0.35\textwidth]{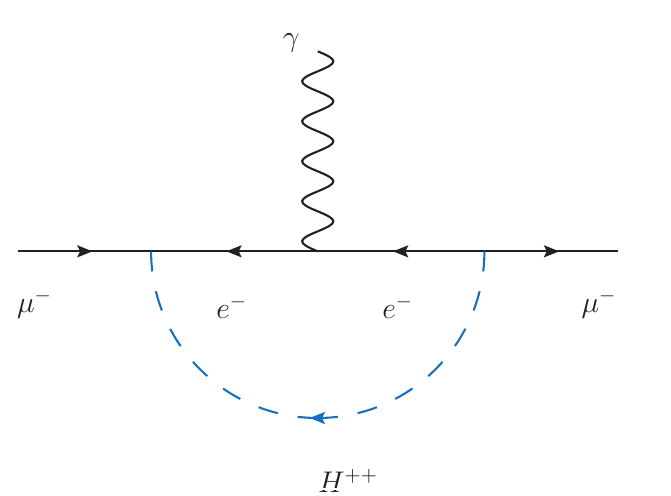}}
    \caption{Feynman diagrams for the $(g-2)_\mu$ in case (b): $H_{3}$, $|Y_{e\mu}| \neq 0$ and case (d): $H^{\pm\pm}$, $|Y_{e\mu}| \neq 0$. Feynman diagrams for the $(g-2)_\mu$ in case (a): $H_{3}$, $|Y_{ee}| = |Y_{\mu\mu}| \neq 0$ and case (c): $H^{\pm\pm}$, $|Y_{ee}| = |Y_{\mu\mu}| \neq 0$ can be obtained simply by changing all the $e$ indices to $\mu$ in the figures. Feynman diagrams for the $(g-2)_e$ can be obtained simply by exchanging all the $e$ and $\mu$ indices in the corresponding $(g-2)_\mu$ diagrams.}
    \label{fig:gminus2}
\end{figure*}

As for the electron~\cite{Parker2018MeasurementOT,morel2020determination} and muon~\cite{Muong-2:2021ojo} anomalous magnetic moments, I have checked and agree with the previous theoretical expressions~\cite{Queiroz:2014zfa,Lindner:2016bgg} for the $\Delta a_e$ and $\Delta a_\mu$ contributions induced by neutral and doubly charged scalar fields. Specifically, the $(g-2)_\mu$ contributions in case (b) and case (d) are:

\begin{subequations}
\begin{align}
    \setcounter{equation}{1}
    \Delta& a_{\mu}^{\rm (b)} = \frac{1}{8\pi^{2}}\frac{m^{2}_{\mu}}{m^{2}_{H_{3}}}\int_{0}^{1}{\rm d}x\frac{|Y_{e\mu}|^{2}x^{2}(1-x+\frac{m_{e}}{m_{\mu}})}{(1-x)(1-\frac{m_{\mu}^{2}}{m_{H_{3}}^{2}}x)+\frac{m_{e}^{2}}{m_{H_{3}}^{2}}x},\\
    \setcounter{equation}{3}
    \Delta& a_{\mu}^{\rm (d)} = -\frac{1}{\pi^{2}}\frac{m^{2}_{\mu}}{m^{2}_{H^{\pm\pm}}}\int_{0}^{1}{\rm d}x\frac{|Y_{e\mu}|^{2}x(1-x)(x+\frac{m_{e}}{m_{\mu}})}{\frac{m_{e}^{2}}{m_{H^{\pm\pm}}^{2}}(1-x)(1-\frac{m_{\mu}^{2}}{m_{e}^{2}}x)+x}\nonumber\\
    &-\frac{1}{2\pi^{2}}\frac{m^{2}_{\mu}}{m^{2}_{H^{\pm\pm}}}\int_{0}^{1}{\rm d}x\frac{|Y_{e\mu}|^{2}x^{2}(1-x+\frac{m_{e}}{m_{\mu}})}{(1-x)(1-\frac{m_{\mu}^{2}}{m_{H^{\pm\pm}}^{2}}x)+\frac{m_{e}^{2}}{m_{H^{\pm\pm}}^{2}}x}.
\end{align}
\label{eq:gminus2}
\end{subequations}
The corresponding Feynman diagrams are shown in Fig.~\ref{fig:gminus2}. The Feynman diagrams and the expressions of the $(g-2)_\mu$ contribution in case (a) and case (c) can be easily obtained by changing all the $e$ indices to $\mu$ in Fig.~\ref{fig:gminus2} and Eq.~\eqref{eq:gminus2}. The Feynman diagrams and the expressions of the $(g-2)_e$ can be easily obtained by exchanging all the $e$ and $\mu$ indices in the corresponding $(g-2)_\mu$ figures and expressions.

\section{Signal and background analysis}
\label{sec:collider}
I focus on the ILC~\cite{Evans:2017rvt} and CLIC~\cite{CLIC:2016zwp} as two benchmark machines for future lepton colliders and present in Table~\ref{tab:colliders} their planned final center-of-mass energy $\sqrt{s}$ and the expected integrated luminosity $\mathscr{L}_{\rm int}$.

\begin{table}[th!]
\centering
    \caption{The planned center-of-mass energy and expected integrated luminosity for the International Linear Collider (ILC) and two stages of Compact Linear Collider (CLIC)}
    \centering
    \begin{tabular}{c|cc}
        \hline
        \hline
         Collider&$\sqrt{s}$ (TeV)&$\mathscr{L}_{\rm int}$ (ab$^{-1}$)\\ \hline
         ILC&1.0&4.0\\ \hline
         \multirow{2}*{CLIC}&
         1.5&2.5\\
         \cline{2-3}
         &3.0&5.0\\ \hline\hline
    \end{tabular}
    \label{tab:colliders}
\end{table}

Based on the four cases mentioned above, I propose two collider signals that can be used to test the $(m_{\rm scalar},Y_{\alpha \beta})$ parameter space: $e^+ e^- \to e^+ e^- \mu^+ \mu^-$ and $e^+ e^- \to e^+ e^+ \mu^- \mu^- / e^- e^- \mu^+ \mu^+$. The second signal that has two same-sign dilepton pairs violates the lepton flavor and the SM background mainly comes from the misidentification of the lepton flavor in the final states. For lepton colliders, the mis-ID rate for electron and muon is less than 0.5\%~\cite{Yu:2017mpx}. This makes $e^\pm e^\pm \mu^\mp \mu^\mp$ almost background-free and I find that the background would not have substantial effects on the estimates of the signal sensitivities. For simplicity, I neglect the $e^\pm e^\pm \mu^\mp \mu^\mp$ SM background for all the prospects below. In this work, I only consider the $ee\mu\mu$ type of the final state. The $eee\mu$ or $e\mu\mu\mu$ type is closely related to the process $\mu \to e\gamma$ and $\mu \to eee$ which is not possible in the cases considered in this paper. The $eeee$ and $\mu\mu\mu\mu$ types are possible in the nonzero diagonal coupling case (a) and (c), and will give a similar cross section compared to their corresponding $ee\mu\mu$ channel.

In this work, I perform a simulation for the signal and background processes using \textsc{MadGraph5\_aMC@NLO}\xspace~\cite{Alwall:2014hca}, requiring all leptons in the final states to satisfy the minimal trigger cuts $p_{\rm T}>10$ GeV, $|\eta|<2.5$, and $\Delta R>0.4$.

The multilepton channels discussed here at future lepton colliders are very clean because we can reconstruct the scalar mass from the dilepton invariant mass. Nevertheless, there are irreducible SM backgrounds, as shown in Fig.~\ref{fig:invariant mass}, which make it difficult to disentangle the signal for smaller Yukawa couplings. This has been taken into account while deriving the sensitivity contours.

In this work, I only present the cross sections at the leading order. But since these are electroweak processes, the NLO corrections are expected to be small, and hence, the $k$-factors should be close to the identity.

\begin{figure*}[t!]
    \centering
    {\includegraphics[width=0.4\textwidth]{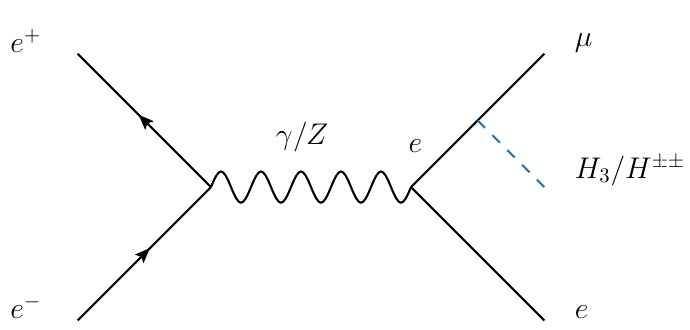}}
    {\includegraphics[width=0.4\textwidth]{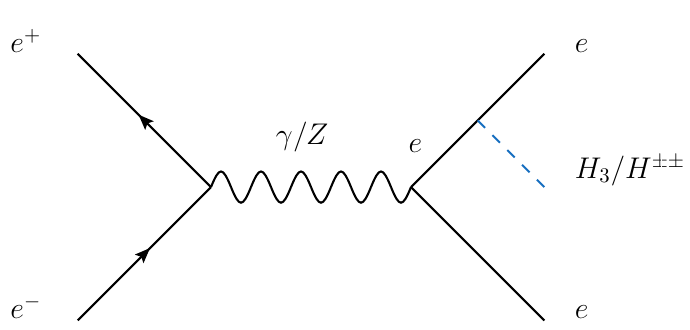}}
    
    {\includegraphics[width=0.4\textwidth]{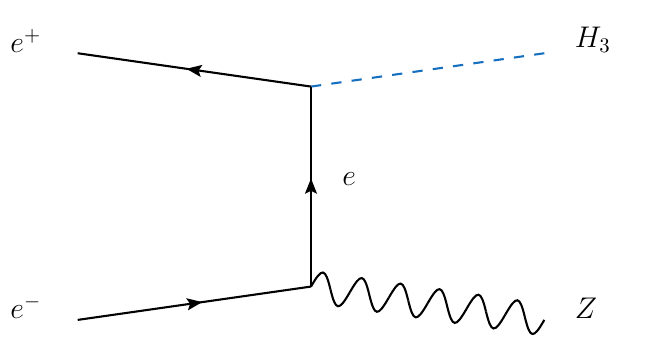}}
    {\includegraphics[width=0.4\textwidth]{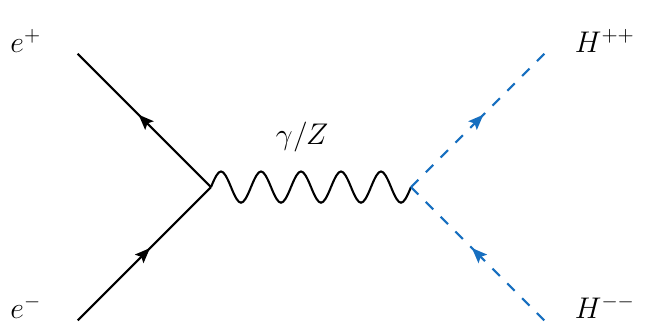}}
    \caption{Representative Feynman diagrams for the single production of $H_{3}$, $H^{\pm\pm}$ and pair production of $H^{\pm\pm}$.}
    \label{fig:signal}
\end{figure*}

The $e^+ e^- \mu^+ \mu^-$ and $e^\pm e^\pm \mu^\mp \mu^\mp$ signals come from the single production of $H_{3}$ or $H^{\pm\pm}$, while there are also contributions from the 
Drell–Yan pair production of $H^{\pm\pm}$ which are dominant when $m_{H^{\pm\pm}} \lesssim \sqrt{s} / 2$ as shown in Fig.~\ref{fig:signal}.

I present in Table~\ref{tab:signals} the potential signal(s) for each case along with their corresponding invariant mass distributions. The emergence of a resonance peak in these distributions would signify the existence of a new neutral or doubly charged Higgs, thereby aiding in the further distinction of the signal from the background. From Table~\ref{tab:signals}, we can see that only case (b) can give both signals while the other cases only have one possible signal. As a demonstration, Fig.~\ref{fig:invariant mass} shows the relevant signal and background invariant mass distributions in the $e^+ e^- \mu^+ \mu^-$ channel. The parameter values used in Fig.~\ref{fig:invariant mass} correspond to the values at the * marks in Fig.~\ref{fig:results}. The clean red peaks in Fig.~\ref{fig:invariant mass} mainly come from the prompt decay of neutral or doubly charged scalar. The invariant mass distributions in the $e^\pm e^\pm \mu^\mp \mu^\mp$ channel are not shown, since they are almost background-free, and the signal distributions are very similar to the ones I present in Fig.~\ref{fig:invariant mass} except for the charge of the electron or muon. In Fig.~\ref{fig:invariant mass}(b1) or (d1), the signal events that mainly range from $0 \sim 500$~GeV correspond to the events in the peak of Fig.~\ref{fig:invariant mass}(b2) or (d2) and vice versa. For example, in Fig.~\ref{fig:invariant mass}(d1), the signal events that ranging from $0 \sim 500$~GeV mainly come from the process $e^+ e^- \to e^+ \mu^+ (H_{\rm L}^{--} \to e^- \mu^-)$ and should peak around 950~GeV in Fig.~\ref{fig:invariant mass}(d2), because in Fig.~\ref{fig:invariant mass}(d2), it shows the invariant mass distribution of $e^- \mu^-$. This is the feature of the single production channel and would be useful to enhance the signal sensitivity as one can choose a selection like $M_{e^+ \mu^-} || M_{e^- \mu^+} > 450$~GeV in Fig.~\ref{fig:invariant mass}(b) or $M_{e^+ \mu^+} || M_{e^- \mu^-} > 900$~GeV in Fig.~\ref{fig:invariant mass}(d).

\begin{table*}[th!]
\centering
\caption{Possible signal(s) for each case and the corresponding invariant mass distributions that could be used to distinguish the signal from the background. ``-'' means the signal is not possible (except for the mis-ID) in the corresponding case.}
\begin{tabular}{l|c|c}
\hline\hline
                                               & $e^+ e^- \to e^+ e^- \mu^+ \mu^-$                                & $e^+ e^- \to e^\pm e^\pm \mu^\mp \mu^\mp$                        \\ \hline
(a) $H_{3},|Y_{ee}| = |Y_{\mu\mu}|$                & $M_{e^+ e^-}\&M_{\mu^+ \mu^-}$ & --                                                        \\ \hline
(b) $H_{3},|Y_{e\mu}|$                             & $M_{e^+ \mu^-}\&M_{e^- \mu^+}$ & $M_{e^+ \mu^-}\&M_{e^- \mu^+}$ \\ \hline
(c) $H^{\pm\pm},|Y_{ee}| = |Y_{\mu\mu}|$ & --                                                        & $M_{e^+ e^+}\&M_{\mu^+ \mu^+}$ \\ \hline
(d) $H^{\pm\pm},|Y_{e\mu}|$              & $M_{e^+ \mu^+}\&M_{e^- \mu^-}$ & --                                                        \\ 
\hline\hline
\end{tabular}
\label{tab:signals}
\end{table*}

Except for the basic cuts mentioned above, I also apply some specific cuts to enhance the sensitivity based on the characteristic signatures in each case:

\begin{enumerate}
    \item [(a)] $H_{3}$ with diagonal couplings: Only $e^+ e^- \mu^+ \mu^-$ final state is possible in this case, where $e^+ e^- \to e^+ e^- (H_3 \to \mu^+ \mu^-)$ or $e^+ e^- \to \mu^+ \mu^- (H_3 \to e^+ e^-)$. Another important channel comes from $e^+ e^- \to ZH_{3} \to e^+ e^- \mu^+ \mu^-$. We expect the dielectron and dimuon invariant mass $M_{e^+e^-}$ and $M_{\mu^+\mu^-}$ to peak at $Z$ and $H_3$ mass; see Figs.~\ref{fig:invariant mass}(a1) and (a2). However, the SM background has similar $M_{e^+e^-}$ and $M_{\mu^+\mu^-}$ distributions around the $Z$ peak. I find that applying cut on $M_{\mu^+\mu^-}$ or $M_{e^+e^-}$ cannot improve the sensitivities much, and make the sensitivities worse in the region where $m_{H_3} \approx m_Z$. So, I first do not apply any further cut for this case. As a comparison, I also show sensitivities with the cut $M_{\mu^+\mu^-} > 120$~GeV in Fig.~\ref{fig:results}(a). The red, yellow, and blue solid (dashed) contours show the 3$\sigma$ sensitivities of the signal without (with) applying the cut $M_{\mu^+\mu^-} > 120$~GeV in Fig.~\ref{fig:results}(a). As we can see, the red, yellow, and blue dashed contours cannot improve the sensitivities much and are not valid when $m_{H_3} \lesssim 120$~GeV.
    \item [(b)] $H_{3}$ with off-diagonal couplings: Both $e^+ e^- \mu^+ \mu^-$ and $e^\pm e^\pm \mu^\mp \mu^\mp$ final states are possible in this case. For the $e^+ e^- \mu^+ \mu^-$ final state, I further apply the cut $M_{e^+e^-},M_{\mu^+\mu^-} > 120$~GeV to reduce the events with a $Z$ boson decaying to a pair of leptons. We also expect the invariant mass $M_{e^\pm \mu^\mp}$ to peak at the $H_{3}$ mass; see Figs.~\ref{fig:invariant mass}(b1) and (b2). Since we do not know the mass of $H_{3}$ and cannot tell where the peak should be around, I do not apply cut on $M_{e^\pm \mu^\mp}$ for this case. Because $e^\pm e^\pm \mu^\mp \mu^\mp$ is almost background-free, I do not apply any further cut in this channel. In Fig.~\ref{fig:results}(b), the red, yellow, and blue solid (dashed) contours now show the 3$\sigma$ sensitivities in $e^+ e^- \mu^+ \mu^-$ ($e^\pm e^\pm \mu^\mp \mu^\mp$) channel. Because the background is small in the $e^\pm e^\pm \mu^\mp \mu^\mp$ channel (assumed to be zero in this work), it is not surprising that the red, yellow, and blue dashed contours behave better than the solid contours.
    \item [(c)] $H^{\pm\pm}$ with diagonal couplings: Only $e^\pm e^\pm \mu^\mp \mu^\mp$ final state is possible in this case. Because $e^\pm e^\pm \mu^\mp \mu^\mp$ is almost background-free, I do not apply any further cut for this case.
    \item [(d)] $H^{\pm\pm}$ with off-diagonal couplings: Only $e^+ e^- \mu^+ \mu^-$ final state is possible in this case, where $e^+ e^- \to e^\mp \mu^\mp (H^{\pm\pm} \to e^\pm \mu^\pm)$. We expect the $e^\pm \mu^\pm$ invariant mass $M_{e^\pm \mu^\pm}$ to peak around the $H^{\pm\pm}$ mass; see Figs.~\ref{fig:invariant mass}(d1) and (d2). Because the pair production channel that is independent of the Yukawa coupling is dominant when $m_{H^{\pm\pm}} \lesssim \sqrt{s} / 2$, the sensitivity in the $(m_{H^{\pm\pm}},Y_{e \mu})$ parameter space is only valid in the region where $m_{H^{\pm\pm}} \gtrsim \sqrt{s} / 2$. Based on this feature, I further apply the cut $M_{e^\pm \mu^\pm} \gtrsim \sqrt{s} / 2$ to maximize the sensitivity. To be specific, I require $M_{e^\pm \mu^\pm} > 500$~GeV at ILC 1.0 TeV stage, $M_{e^\pm \mu^\pm} > 750$~GeV at CLIC 1.5 TeV stage, and $M_{e^\pm \mu^\pm} > 1400$~GeV at CLIC 3.0 TeV stage.
\end{enumerate}

I summarize the further selections used in my analysis in Table~\ref{tab:selection} in the $e^+ e^- \mu^+ \mu^-$ channel.

\begin{table}[ht!]
\centering
\caption{Further selections for the analysis for each case in the $e^+ e^- \mu^+ \mu^-$ channel}
\begin{tabular}{c|c}
\hline\hline
$e^+ e^- \mu^+ \mu^-$            & Selection                                        \\ \hline
case (a)   & -- or $M_{\mu^+ \mu^-} > 120~$GeV            \\ \hline
case (b)   & $M_{e^+ e^-},M_{\mu^+ \mu^-} > 120~$GeV     \\ \hline
case (d)   & $M_{e^+ \mu^+},M_{e^- \mu^-} \gtrsim \sqrt{s}/2$ \\ \hline\hline
\end{tabular}
\label{tab:selection}
\end{table}

\section{Results} \label{sec:results}
\begin{figure*}[htbp!]
    \centering
    \subfloat[$m_{H_3}$, $|Y_{ee}| = |Y_{\mu\mu}| \neq 0$]{\includegraphics[width=0.5\textwidth]{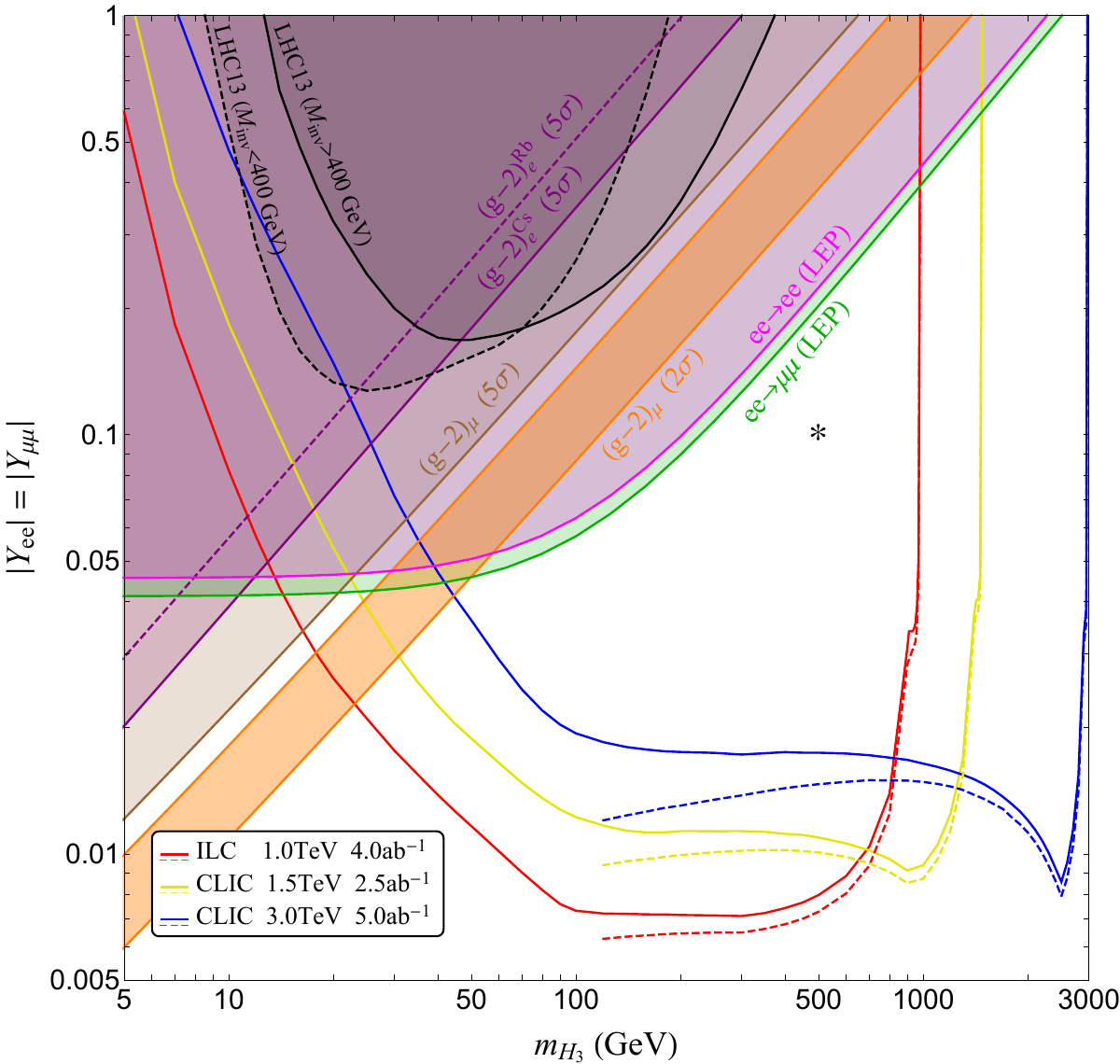}}
    \subfloat[$m_{H_3}$, $|Y_{e\mu}| \neq 0$]{\includegraphics[width=0.5\textwidth]{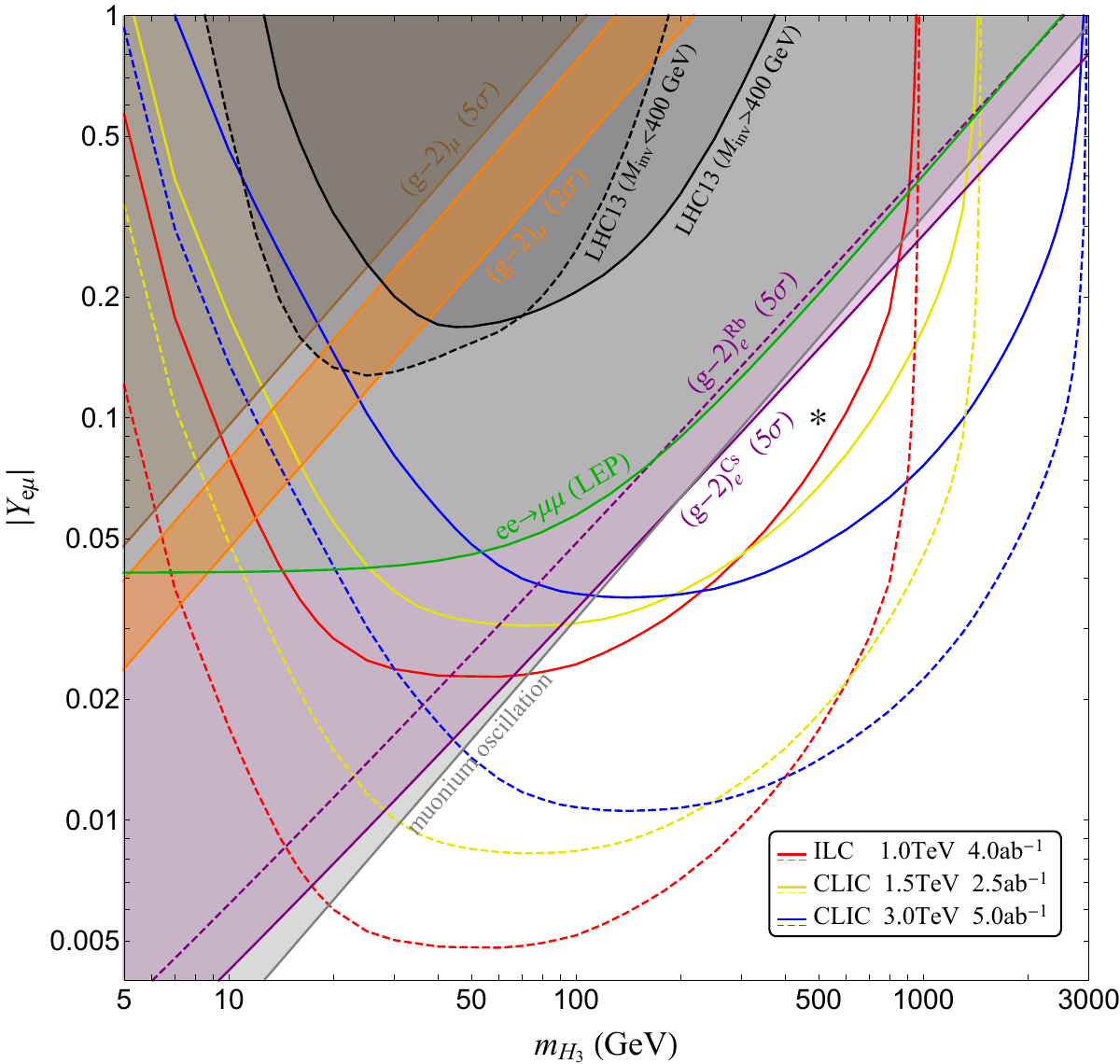}}\\
    \subfloat[$m_{H_{\rm L,R}^{\pm\pm}}$, $|Y_{ee}| = |Y_{\mu\mu}| \neq 0$]{\includegraphics[width=0.5\textwidth]{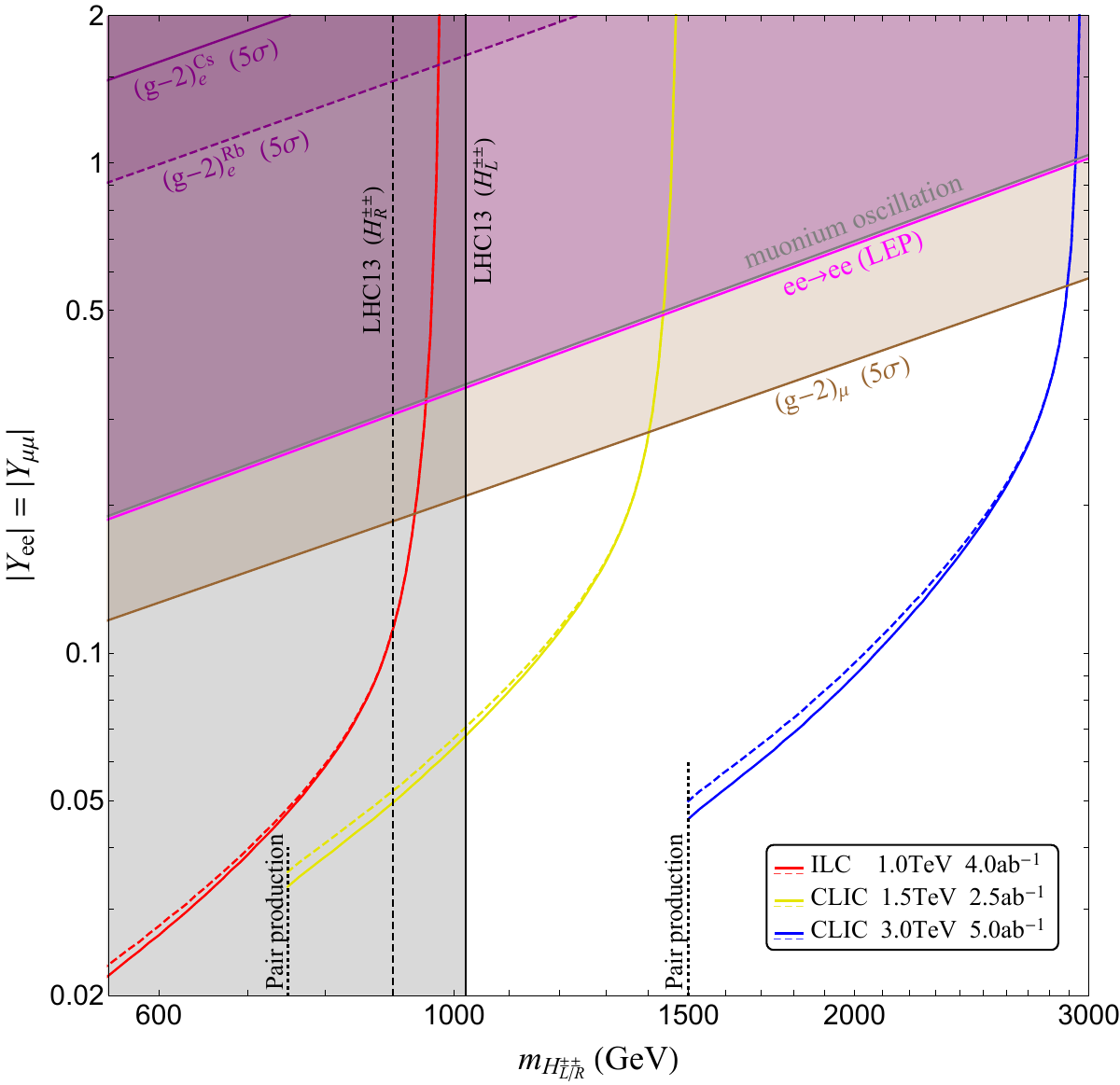}}
    \subfloat[$m_{H_{\rm L,R}^{\pm\pm}}$, $|Y_{e\mu}| \neq 0$]{\includegraphics[width=0.5\textwidth]{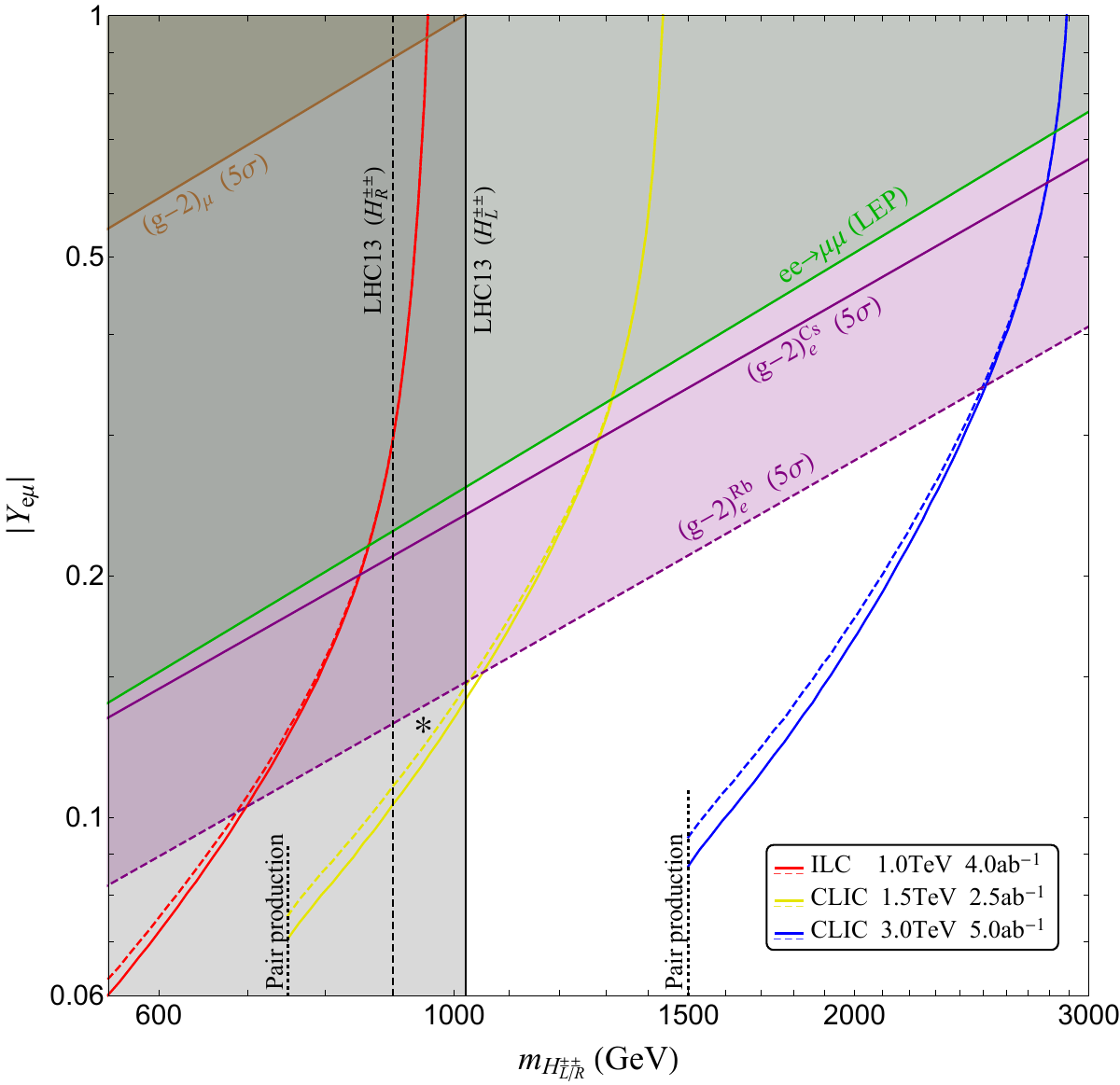}}
    \caption{Four cases in the $H_{3}$ or $H_{\rm L,R}^{\pm\pm}$ parameter space. The orange-shaded regions explain the $(g-2)_\mu$ anomaly at $2\sigma$ CL, while the brown-shaded regions display the $5\sigma$ disfavored regions of $\Delta a_\mu$. The purple-shaded regions with solid (dashed) boundaries are the $5\sigma$ disfavored regions of $\Delta a_e$ using Cs (Rb) measurements. The black solid and dashed contours are the 13~TeV limits derived from the different LHC multilepton searches discussed in the text. Other shaded regions show the relevant constraints on the parameter space from muonium oscillation (gray), LEP $ee \to ee$ (magenta), and LEP $ee \to \mu\mu$ (green). The red, yellow, and blue contours respectively show the $3\sigma$ sensitivities of the 1.0~TeV ILC, 1.5~TeV CLIC, and 3.0~TeV CLIC in the (a) $e^+ e^- \mu^+ \mu^-$ channel without (solid contours) or with (dashed contours) $M_{\mu^+ \mu^-} > 120$~GeV; (b) $e^+ e^- \mu^+ \mu^-$ channel (solid contours) and $e^\pm e^\pm \mu^\mp \mu^\mp$ channel (dashed contours); (c) $e^\pm e^\pm \mu^\mp \mu^\mp$ channel (solid contours for $H_{\rm L}^{\pm\pm}$ and dashed contours for $H_{\rm R}^{\pm\pm}$); (d) $e^+ e^- \mu^+ \mu^-$ channel (solid contours for $H_{\rm L}^{\pm\pm}$ and dashed contours for $H_{\rm R}^{\pm\pm}$). The * marks in the figures give the representative values of the corresponding parameters used in Fig.~\ref{fig:invariant mass}.}
    \label{fig:results}
\end{figure*}

In the Left-Right Symmetric Model~\cite{Pati:1974yy,Mohapatra:1974gc,Senjanovic:1975rk,Mohapatra:1977mj,Cai:2017mow,BhupalDev:2018vpr,BhupalDev:2018tox}, the triplet Higgs fields $\Delta_{\rm L}$ and $\Delta_{\rm R}$ transform as triplets under $SU(2)_L$ and $SU(2)_R$ gauge symmetries respectively. Both of them give rise to a doubly charged Higgs. They are named as $H^{\pm\pm}_{\rm L}$ and $H^{\pm\pm}_{\rm R}$. Besides, the doubly charged Higgs in the canonical type-II seesaw model~\cite{Akeroyd:2005gt,Hektor:2007uu,Chao:2008mq} is the same as $H^{\pm\pm}_{\rm L}$ and the doubly charged scalar in the Zee–Babu neutrino mass model~\cite{Zee:1985rj,Zee:1985id,Babu:1988ki} has the same quantum numbers as $H^{\pm\pm}_{\rm R}$\footnote{I assume the $Z'$ is much heavier, then the electroweak production of the doubly charged scalar in the Zee–Babu model is the same as $H^{\pm\pm}_{\rm R}$.}. In case (c) and case (d) of this work, I consider both $H^{\pm\pm}_{\rm L}$ and $H^{\pm\pm}_{\rm R}$. Because they have different couplings to the $Z$ boson~\cite{Fuks:2019clu}, their sensitivities (the red, yellow, and blue solid or dashed contours in Figs.~\ref{fig:results}(c) and (d)) are a little bit different but the LHC constraints derived from the Drell–Yan pair production process (the black solid or dashed contours in Figs.~\ref{fig:results}(c) and (d)) have a sizable difference\footnote{The cross section for the pair production of $H^{\pm\pm}_{\rm L}$ is roughly two times larger than the $H^{\pm\pm}_{\rm R}$~\cite{ATLAS:2022pbd}, which makes the constraints of $H^{\pm\pm}_{\rm L}$ stronger than $H^{\pm\pm}_{\rm R}$ in Figs.~\ref{fig:results}(c) and (d).}.

I consider all the four cases where $H_{3}$ or $H^{\pm\pm}_{\rm L,R}$ has nonzero diagonal or off-diagonal Yukawa couplings in the $e,\mu$ sector. In Fig.~\ref{fig:results}, the orange-shaded regions explain the $(g-2)_\mu$ anomaly~\cite{Muong-2:2021ojo} at $2\sigma$ CL, while the brown-shaded regions display the $5\sigma$ disfavored regions of $\Delta a_\mu$. As shown in Eq.~\eqref{eq:gminus2}, the doubly charged scalar has an opposite contribution to $\Delta a_\mu$, so there is no $2\sigma$ favored region in Figs.~\ref{fig:results}(c) and (d). The purple-shaded regions in Fig.~\ref{fig:results} with solid (dashed) boundaries are the $5\sigma$ disfavored regions of $\Delta a_e$ using Cs~\cite{Parker2018MeasurementOT} (Rb~\cite{morel2020determination}) measurements. Again, because the doubly charged scalar has an opposite contribution to $\Delta a_e$ compared with the neutral scalar, the $(g-2)_e^{\rm Cs}$ constraint is stronger in the $H_{3}$ parameter space (Figs.~\ref{fig:results}(a) and (b)), while the $(g-2)_e^{\rm Rb}$ constraint is stronger in the $H^{\pm\pm}_{\rm L,R}$ parameter space (Figs.~\ref{fig:results}(c) and (d)). I recast the ATLAS multilepton analysis~\cite{ATLAS-CONF-2021-011} using the Signal Region $4\ell$ Off-Z with $M_{\rm inv} > (<) 400$~GeV and set new bounds on the neutral scalar $H_{3}$ shown as the black solid (dashed) contours in Figs.~\ref{fig:results}(a) and (b). The black solid (dashed) contours in Figs.~\ref{fig:results}(c) and (d) are the 95\% CL limits on $m_{H_{\rm L}^{\pm\pm}}$ ($m_{H_{\rm R}^{\pm\pm}}$) from the LHC multilepton search~\cite{ATLAS:2022pbd}, assuming $\Sigma_{\ell\ell^\prime}{\rm BR}(H^{\pm\pm} \to \ell^\pm \ell^{\prime\pm}) = 100\%$. Other shaded regions in Fig.~\ref{fig:results} show the relevant constraints on the parameter space from muonium oscillation~\cite{Willmann:1998gd} (gray), LEP $ee \to ee$~\cite{DELPHI:2005wxt} (magenta), and LEP $ee \to \mu\mu$~\cite{DELPHI:2005wxt} (green).

The red, yellow, and blue contours in Fig.~\ref{fig:results} show the $3\sigma$ sensitivities of the 1.0~TeV ILC, 1.5~TeV CLIC, and 3.0~TeV CLIC in the $ee\mu\mu$ channels respectively. To be specific, in Fig.~\ref{fig:results}(a), $H_{3}$ with nonzero diagonal Yukawa couplings, the red, yellow, and blue solid (dashed) contours show the $3\sigma$ sensitivities in the $e^+ e^- \mu^+ \mu^-$ channel without (with) cut $M_{\mu^+\mu^-} > 120$~GeV; in Fig.~\ref{fig:results}(b), $H_{3}$ with nonzero off-diagonal Yukawa couplings, the red, yellow, and blue solid (dashed) contours show the $3\sigma$ sensitivities in the $e^+ e^- \mu^+ \mu^-$ ($e^\pm e^\pm \mu^\mp \mu^\mp$) channel; in Fig.~\ref{fig:results}(c), $H_{\rm L}^{\pm\pm}$ ($H_{\rm R}^{\pm\pm}$) with nonzero diagonal Yukawa couplings, the red, yellow, and blue solid (dashed) contours show the $3\sigma$ sensitivities in the $e^\pm e^\pm \mu^\mp \mu^\mp$ channel; in Fig.~\ref{fig:results}(d), $H_{\rm L}^{\pm\pm}$ ($H_{\rm R}^{\pm\pm}$) with nonzero off-diagonal Yukawa couplings, the red, yellow, and blue solid (dashed) contours show the $3\sigma$ sensitivities in the $e^+ e^- \mu^+ \mu^-$ channel.

In Fig.~\ref{fig:results}, all the red, yellow, and blue contours asymptotically approach the line $m_{H_{3},H_{\rm L,R}^{\pm\pm}} = \sqrt{s}$ because this is the search of the single production channels of the $H_{3}$ and $H_{\rm L,R}^{\pm\pm}$. In case (a), because there is another important channel for the $H_3$ single production: $e^+ e^- \to Z H_3 \to e^+ e^- \mu^+ \mu^-$, the red, yellow, and blue contours in Fig.~\ref{fig:results}(a) have a kink at $m_{H_3} \approx \sqrt{s} - 90$~GeV which displays the feature that an on-shell $Z$ boson turning into an off-shell $Z$ boson as we increase the mass of $H_3$.

The red, yellow, blue, and black curves in Figs.~\ref{fig:results}(a) and (b) are increasing fast as the mass of $H_3$ decrease in the low-mass region because the leptons in the final state are soft and cannot pass the selection of the corresponding $p_{\rm T}$ cutoff. This feature means that, although the on-shell production of $H_3$ could be very large, the ability to detect the on-shell, low-mass $H_3$ at colliders is still not promising. However, although the on-shell searches are not sensitive in the low-mass range because of soft leptons, the LEP constraints in Figs.~\ref{fig:results}(a) and (b) give a flat (this is because the EFT approach for the LEP $e e \to \ell \ell$ data is only sensitive to the couplings) and stronger limit when $m_{H_3} \lesssim 10$~GeV.


\section{Discussions}
\label{sec:discussions}

\indent $\bullet$ {\bf Other possible bounds:}
For the neutral Higgs cases, the future lepton collider constraints I proposed depend on the overall Yukawa couplings while the current constraints mostly come from low-energy data, such as $B$ and $K$ meson mixing~\cite{Dev:2017dui}. The LHCb $B \to K \mu^+ \mu^-$ limitations are applicable exclusively below 5~GeV~\cite{Dev:2017dui}. These constraints on the mass of $H_3$ depend on the specific model-dependent parameters. The future lepton collider constraints derived here are complementary to the low-energy constraints and extend to higher $H_3$ masses. There also exist astrophysical and cosmological constraints at lower masses ($m_{H_3} < {\cal O}({\rm GeV})$)~\cite{Ibe:2021fed,Balaji:2022noj}. For this work, additional astrophysical and cosmological constraints at lower masses might apply from dark matter direct detection experiments~\cite{Feng:2022inv}, but this requires the Higgs to couple to dark matter.

As for the doubly charged Higgs, there are also studies on the future lepton colliders~\cite{Crivellin:2018ahj,Gluza:2020qrt}, HL-LHC~\cite{Bambhaniya:2015wna,Crivellin:2018ahj,Gluza:2020qrt,CMS-PAS-FTR-22-006,Ruiz:2022sct}, and FCC-hh~\cite{Bambhaniya:2015wna,Gluza:2020qrt} that will work up to $\sqrt{s} = 100$~TeV~\cite{FCC:2018vvp}. At HL-LHC, the Drell-Yan-like pair production channel can improve the limit of the doubly charged Higgs to $\sim$1400~GeV at 95\% CL~\cite{CMS-PAS-FTR-22-006}.

In this work, I only show the constraints which depend on the relevant Yukawa couplings and masses. However, there might be additional constraints in a specific model like HTM or LRSM~\cite{Gluza:2020qrt}, which are stronger. For instance, the Møller scattering limit is stronger than the LEP limit in the parity-violating LRSM~\cite{Dev:2018sel,Gluza:2020qrt}. Similarly, the $\rho$-parameter constraint applies to the HTM with large triplet VEV~\cite{Gluza:2020qrt}.

$\bullet$ {\bf Electron and muon (g-2):}
Remarkably, we observe that in Figs.~\ref{fig:results}(a) and (c), the $(g-2)_\mu$ bounds are more stringent than the $(g-2)_e$ ones, whereas in Figs.~\ref{fig:results}(b) and (d), the $(g-2)_e$ bounds are stronger than the $(g-2)_\mu$ ones. Roughly speaking, for the off-diagonal Yukawa coupling case in Eq.~\eqref{eq:gminus2}, the dominant contribution to $(g-2)_\mu$ comes from $\Delta a_\mu^{\rm off-diagonal} \propto \frac{m_\mu^2}{m_H^2} \int_{0}^{1}{\rm d}x \ x^2 = \frac{m_\mu^2}{3m_H^2}$, while the dominant contribution to $(g-2)_e$ should be $\Delta a_e^{\rm off-diagonal} \propto \frac{m_e^2}{m_H^2} \int_{0}^{1}{\rm d}x \ \frac{x^2(m_\mu / m_e)}{1-x+m_\mu^2 / m_H^2} \approx -\frac{m_e m_\mu}{m_H^2}{\rm ln}(\frac{m_\mu^2}{m_H^2})$. This gives $\left(\frac{\Delta a_e}{\Delta a_\mu}\right)_{\rm off-diagonal} \sim -3\frac{m_e}{m_\mu}{\rm ln}(\frac{m_\mu^2}{m_H^2}) \sim {\cal O}(0.01)$ to ${\cal O}(0.1)$. But for the diagonal Yukawa coupling case, the ratio is simply $\left(\frac{\Delta a_e}{\Delta a_\mu}\right)_{\rm diagonal} \sim \frac{m_e^2}{m_\mu^2} \sim {\cal O}(10^{-5})$. As a result, the $(g-2)_e$ gets a relatively larger contribution in the off-diagonal coupling case than the diagonal coupling case compared with the $(g-2)_\mu$. This means the constraints of the $(g-2)_e$ should be stronger in the off-diagonal case. And that is the reason in Figs.~\ref{fig:results}(b) and (d), the $(g-2)_e$ constraints are stronger than the $(g-2)_\mu$ ones, but in Figs.~\ref{fig:results}(a) and (c), the $(g-2)_\mu$ constraints are stronger.

$\bullet$ {\bf New lattice results on muon (g-2):}
In this paper, my goal is not to address the $(g-2)_\mu$ anomaly but rather focus on the discovery prospect of the neutral and doubly charged scalars at future lepton colliders. However, in Fig.~\ref{fig:results}(a), there does exist a parameter space of neutral scalar with its mass ranging from $5 \sim 50$~GeV that can explain $(g-2)_\mu$ and can be partly tested at CLIC in the $e^+ e^- \mu^+ \mu^-$ channel.

The discrepancy of $\Delta a_\mu$ used in this paper comes from the result of the Fermilab Muon $(g-2)$ experiment~\cite{Muong-2:2021ojo}, which is compared with the world-average of the SM prediction using the ``R-ratio method''~\cite{Aoyama:2020ynm} and give a discrepancy of $4.2\sigma$:
\begin{align}
    \Delta a_\mu \equiv a_\mu^{\rm exp} - a_\mu^{\rm SM} = (251 \pm 59) \times 10^{-11} \, .
    \label{eq:deltamu}
\end{align}
But the lattice simulation result from the BMW Collaboration~\cite{Borsanyi:2020mff} increases the leading hadronic contribution of $a_\mu^{\rm SM}$ with a relatively larger uncertainty. There are several new lattice results available now come from other collaborations~\cite{Ce:2022kxy,Alexandrou:2022amy,Colangelo:2022vok,talkLehner2022,talkGottlieb2022,talkColangelo2022} seem to agree with the BMW result and would result in a discrepancy of $\sim 3.3 \sigma$. Note that the center value of $\Delta a_\mu$ determines the position of the orange strip in Fig.~\ref{fig:results}(a) or (b), while the error of $\Delta a_\mu$ determines the width of the orange strip. If the center value of $\Delta a_\mu$ is reduced according to the lattice results and the error is not changed much, the $(g-2)_\mu$ $5\sigma$ constraints in Fig.~\ref{fig:results} would become stronger while the survived $(g-2)_\mu$ $2\sigma$ favored region in Fig.~\ref{fig:results}(a) would shift downwards and become larger (in the direction of length) accordingly, because, for a same value of mass, a smaller coupling would be enough to generate the needed value of $\Delta a_\mu$. As a comparison, the lower orange boundary of the $(g-2)_\mu$ $2\sigma$ favored region in Fig.~\ref{fig:results}(a) or (b) corresponds to $\Delta a_\mu = 133 \times 10^{-11}$, which is about 30\% larger than the BMW center value $\Delta a_\mu = 107 \times 10^{-11}$.

\begin{figure*}[ht!]
    \centering
    \includegraphics[width=0.4\textwidth]{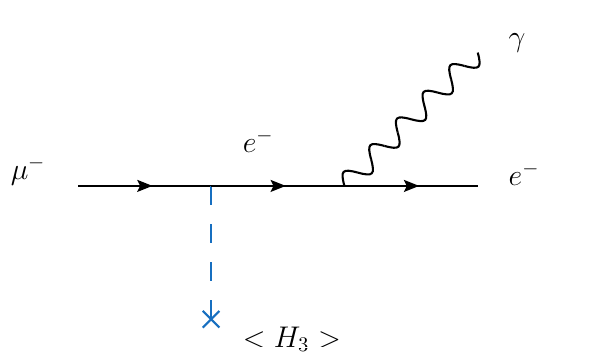}
    \includegraphics[width=0.4\textwidth]{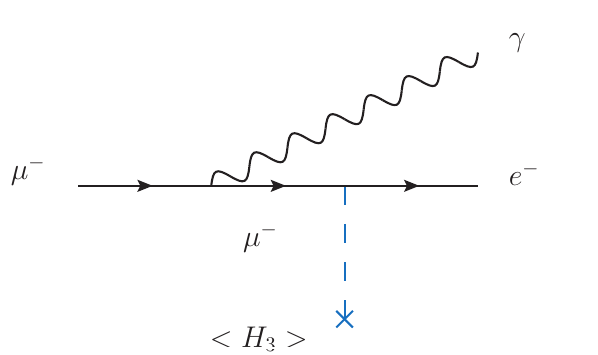}
    \caption{$\mu \to e \gamma$ contributions in case (b) $H_{3}$, $Y_{e\mu} \neq 0$. However, the total amplitude square is zero.}
    \label{fig:mu_to_egamma}
\end{figure*}

$\bullet$ {\bf $\bf{\mu \to e \gamma}$:}
Assuming the diagonal and off-diagonal terms of the Yukawa couplings are not zero separately is crucial. It allows us not to worry about the strong constraint from $\mu \to e \gamma$. One would argue that in case (b), even if only the $Y_{e\mu}$ coupling of $H_{3}$ is not 0, there are still contributions from the VEV of $H_3$; see Fig.~\ref{fig:mu_to_egamma}. Indeed, the spin averaged amplitude square of each diagram in Fig.~\ref{fig:mu_to_egamma} is
\begin{equation}
    \left\langle|M_1|^2\right\rangle \ = \ \left\langle|M_2|^2\right\rangle \ = \ 4e^{2}|Y_{e\mu}|^{2}v^{2}\frac{p_{e}\cdot p_{\mu}-2m_{e}m_{\mu}}{(m_{e}-m_{\mu})^{2}},
\end{equation}
and could be large. But the interference terms will cancel this identically:
\begin{equation}
    \left\langle M_1 M_2^*\right\rangle \ = \ \left\langle M_2 M_1^*\right\rangle \ = \ -4e^{2}|Y_{e\mu}|^{2}v^{2}\frac{p_{e}\cdot p_{\mu}-2m_{e}m_{\mu}}{(m_{e}-m_{\mu})^{2}}.
\end{equation}

\section{Conclusions} \label{sec:conclusions}

Lepton colliders allow for precise measurements of new physics beyond the SM. They produce cleaner collision events with less background noise compared with the hadron colliders.

Focusing on the $e,\mu$ sector of the Yukawa coupling matrix of the neutral and doubly charged scalars, I proposed four characteristic cases and analyzed their discovery prospect in the $e^+ e^- \mu^+ \mu^-$ and $e^\pm e^\pm \mu^\mp \mu^\mp$ channels at future lepton colliders in a model-independent way. I recast the current ATLAS multilepton analysis~\cite{ATLAS-CONF-2021-011} and set new bounds on the neutral scalar $H_{3}$. I also made a detailed investigation, outlining various di-lepton invariant mass distributions in discriminating signals in each of the cases from backgrounds and from each other. The corresponding Yukawa couplings can be detected ranging from $0.005 \sim 0.5$ at future lepton colliders depending on the cases, $\sqrt{s}$, luminosity, and the mass of the scalar.

I also checked the previous expressions of the electron and muon $(g-2)$ induced by neutral and doubly charged scalar fields~\cite{Queiroz:2014zfa,Lindner:2016bgg}. I further showed that, for both neutral and doubly charged scalar cases, the $(g-2)_e$ gets a relatively larger contribution in the off-diagonal Yukawa coupling cases while the $(g-2)_\mu$ gets a relatively larger contribution in the diagonal Yukawa coupling cases.

\acknowledgments
I am very grateful to Bhupal Dev and Yongchao Zhang for their useful discussions and collaboration during the early stage of this project. I thank Amarjit Soni for the information on the new lattice results of muon $(g-2)$. I also thank Ahmed Ismail for an important remark on $\mu \to e \gamma$ with Fig.~\ref{fig:mu_to_egamma}. This work is supported by the US Department of Energy under Grant No.~DE-SC0017987.

\appendix
\section{Kinematic Distributions}
\label{sec:invariant mass}

\begin{figure*}[ht!]
    \centering
    \setcounter{subfigure}{0}\renewcommand{\thesubfigure}{a\arabic{subfigure}}\makeatletter
    \subfloat[$m_{H_{3}} = 500$~GeV, $|Y_{ee}| = |Y_{\mu\mu}| = 0.1$]{\includegraphics[width=0.45\textwidth]{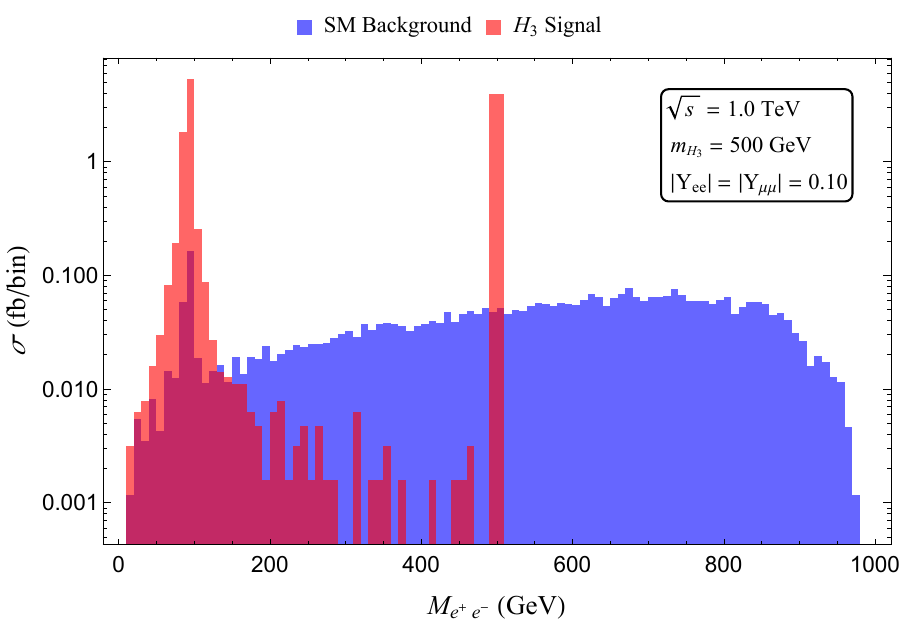}}
    \subfloat[$m_{H_{3}} = 500$~GeV, $|Y_{ee}| = |Y_{\mu\mu}| = 0.1$]{\includegraphics[width=0.45\textwidth]{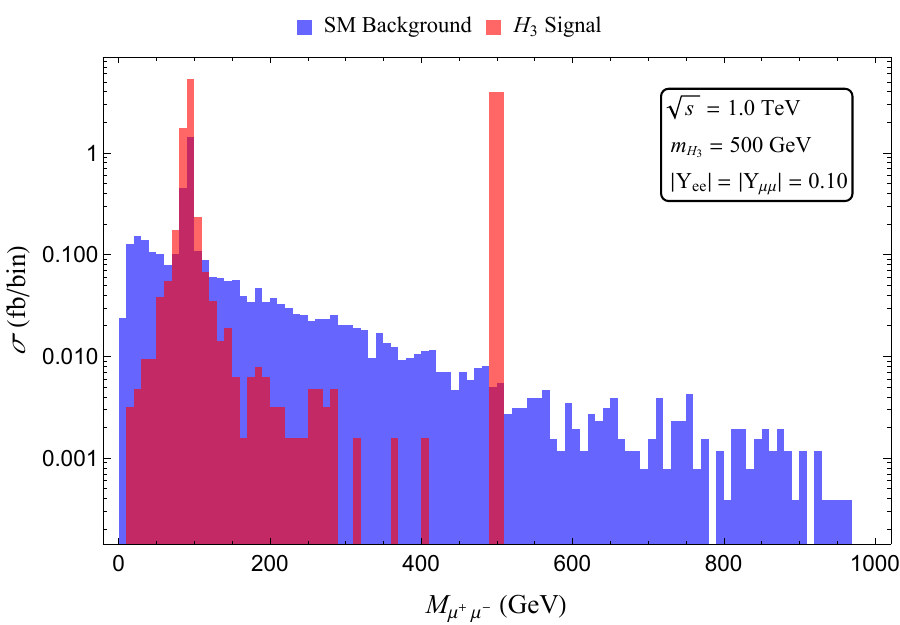}}\\
    \setcounter{subfigure}{0}\renewcommand{\thesubfigure}{b\arabic{subfigure}}\makeatletter
    \subfloat[$m_{H_{3}} = 500$~GeV, $|Y_{e\mu}| = 0.1$]{\includegraphics[width=0.45\textwidth]{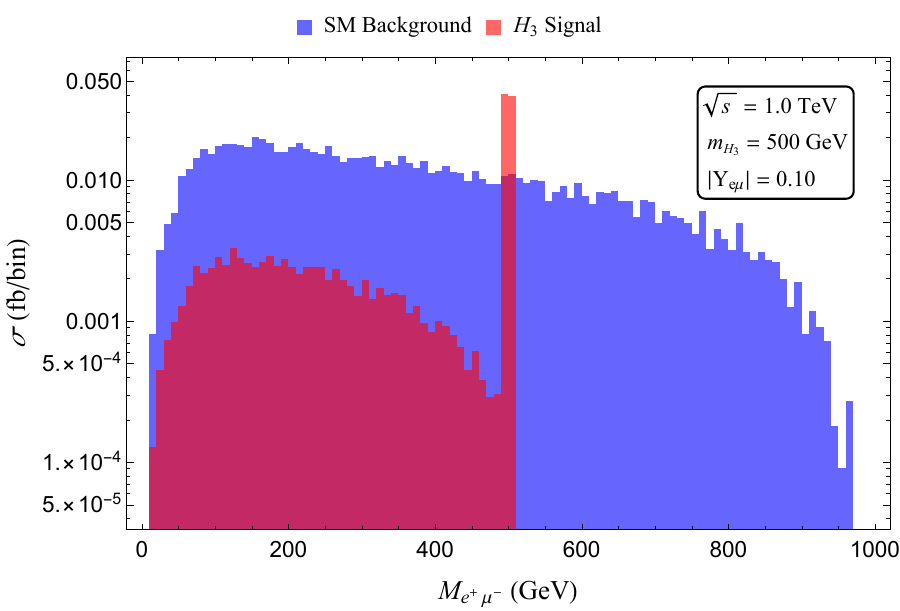}}
    \subfloat[$m_{H_{3}} = 500$~GeV, $|Y_{e\mu}| = 0.1$]{\includegraphics[width=0.45\textwidth]{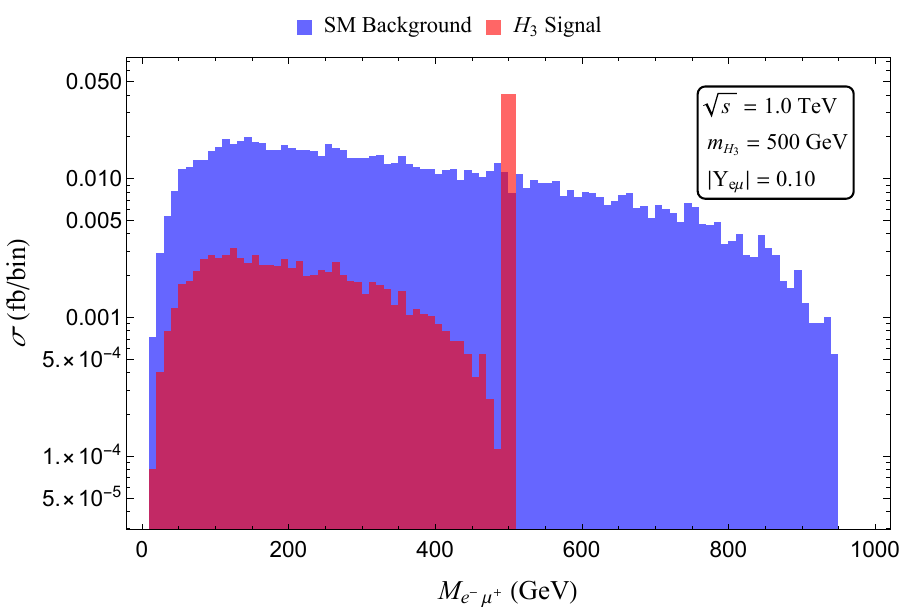}}\\
    \setcounter{subfigure}{0}\renewcommand{\thesubfigure}{d\arabic{subfigure}}\makeatletter
    \subfloat[$m_{H^{\pm\pm}_{\rm R}} = 950$~GeV, $|Y_{e\mu}| = 0.13$]{\includegraphics[width=0.45\textwidth]{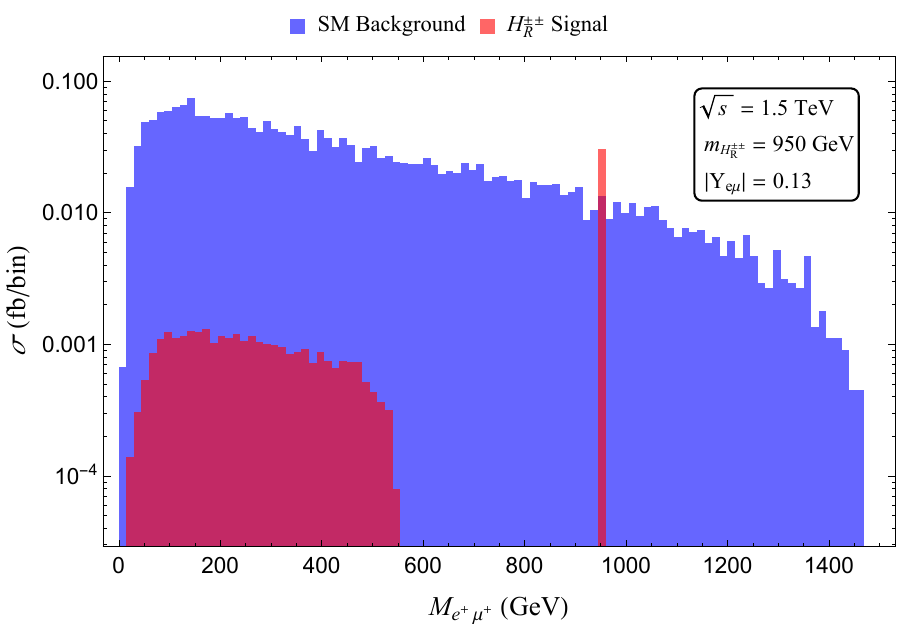}}
    \subfloat[$m_{H^{\pm\pm}_{\rm R}} = 950$~GeV, $|Y_{e\mu}| = 0.13$]{\includegraphics[width=0.45\textwidth]{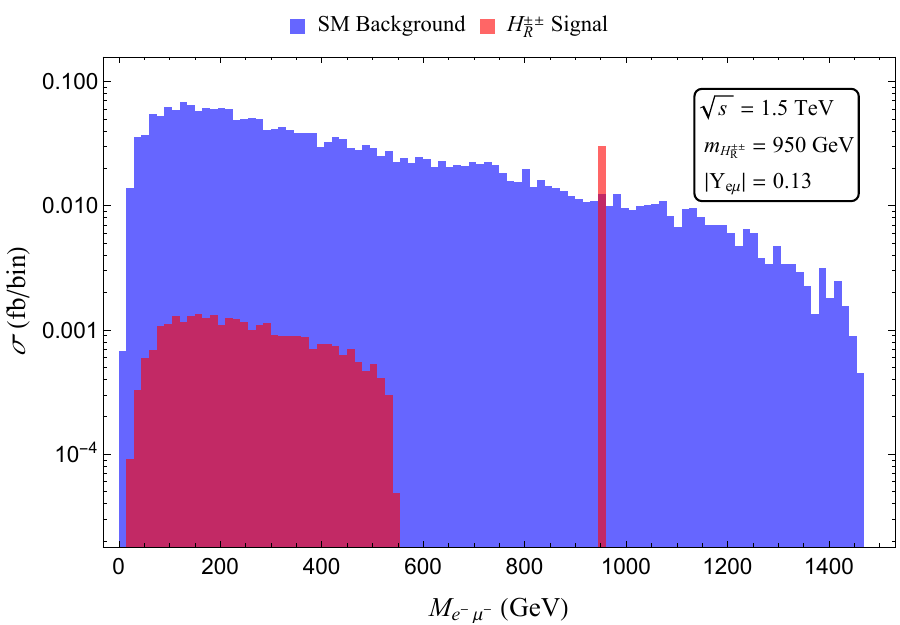}}
    \caption{Invariant mass distributions for case (a) $m_{H_{3}} = 500$~GeV, $|Y_{ee}| = |Y_{\mu\mu}| = 0.1$ at $\sqrt{s} = 1.0$~TeV; (b) $m_{H_{3}} = 500$~GeV, $|Y_{e\mu}| = 0.1$ at $\sqrt{s} = 1.0$~TeV; (d) $m_{H^{\pm\pm}_{\rm L}} = 950$~GeV, $|Y_{e\mu}| = 0.13$ at $\sqrt{s} = 1.5$~TeV signals (red) and SM background (blue) in the $e^+ e^- \to e^+ e^- \mu^+ \mu^-$ channel. The parameter values used here correspond to the values at the * marks in Fig.~\ref{fig:results}.}
    \label{fig:invariant mass}
\end{figure*}

As a demonstration, in Fig.~\ref{fig:invariant mass}, I show the signal and background invariant mass distributions in the $e^+ e^- \mu^+ \mu^-$ channel using the parameter values at the * marks in Fig.~\ref{fig:results}. One can see the red signal resonance peak around the mass of the assumed neutral or doubly charged scalar. But it is not always useful to put a cut on these distributions because the position of the peak is not known at first. However, the signal peaks in Fig.~\ref{fig:invariant mass} would provide direct evidence of the new particle with its mass and charge.

\bibliographystyle{apsrev4-1}
\bibliography{mybib2}

\begin{thebibliography}{66}%
\makeatletter
\providecommand \@ifxundefined [1]{%
 \@ifx{#1\undefined}
}%
\providecommand \@ifnum [1]{%
 \ifnum #1\expandafter \@firstoftwo
 \else \expandafter \@secondoftwo
 \fi
}%
\providecommand \@ifx [1]{%
 \ifx #1\expandafter \@firstoftwo
 \else \expandafter \@secondoftwo
 \fi
}%
\providecommand \natexlab [1]{#1}%
\providecommand \enquote  [1]{``#1''}%
\providecommand \bibnamefont  [1]{#1}%
\providecommand \bibfnamefont [1]{#1}%
\providecommand \citenamefont [1]{#1}%
\providecommand \href@noop [0]{\@secondoftwo}%
\providecommand \href [0]{\begingroup \@sanitize@url \@href}%
\providecommand \@href[1]{\@@startlink{#1}\@@href}%
\providecommand \@@href[1]{\endgroup#1\@@endlink}%
\providecommand \@sanitize@url [0]{\catcode `\\12\catcode `\$12\catcode
  `\&12\catcode `\#12\catcode `\^12\catcode `\_12\catcode `\%12\relax}%
\providecommand \@@startlink[1]{}%
\providecommand \@@endlink[0]{}%
\providecommand \url  [0]{\begingroup\@sanitize@url \@url }%
\providecommand \@url [1]{\endgroup\@href {#1}{\urlprefix }}%
\providecommand \urlprefix  [0]{URL }%
\providecommand \Eprint [0]{\href }%
\providecommand \doibase [0]{http://dx.doi.org/}%
\providecommand \selectlanguage [0]{\@gobble}%
\providecommand \bibinfo  [0]{\@secondoftwo}%
\providecommand \bibfield  [0]{\@secondoftwo}%
\providecommand \translation [1]{[#1]}%
\providecommand \BibitemOpen [0]{}%
\providecommand \bibitemStop [0]{}%
\providecommand \bibitemNoStop [0]{.\EOS\space}%
\providecommand \EOS [0]{\spacefactor3000\relax}%
\providecommand \BibitemShut  [1]{\csname bibitem#1\endcsname}%
\let\auto@bib@innerbib\@empty
\bibitem [{\citenamefont {Gunion}\ \emph {et~al.}(2000)\citenamefont {Gunion},
  \citenamefont {Haber}, \citenamefont {Kane},\ and\ \citenamefont
  {Dawson}}]{Gunion:1989we}%
  \BibitemOpen
  \bibfield  {author} {\bibinfo {author} {\bibfnamefont {J.~F.}\ \bibnamefont
  {Gunion}}, \bibinfo {author} {\bibfnamefont {H.~E.}\ \bibnamefont {Haber}},
  \bibinfo {author} {\bibfnamefont {G.~L.}\ \bibnamefont {Kane}}, \ and\
  \bibinfo {author} {\bibfnamefont {S.}~\bibnamefont {Dawson}},\ }\href@noop {}
  {\emph {\bibinfo {title} {{The Higgs Hunter's Guide}}}},\ Vol.~\bibinfo
  {volume} {80}\ (\bibinfo {year} {2000})\BibitemShut {NoStop}%
\bibitem [{\citenamefont {Gunion}(2002)}]{Gunion:2002in}%
  \BibitemOpen
  \bibfield  {author} {\bibinfo {author} {\bibfnamefont {J.~F.}\ \bibnamefont
  {Gunion}},\ }in\ \href@noop {} {\emph {\bibinfo {booktitle} {{10th
  International Conference on Supersymmetry and Unification of Fundamental
  Interactions (SUSY02)}}}}\ (\bibinfo {year} {2002})\ pp.\ \bibinfo {pages}
  {80--103},\ \Eprint {http://arxiv.org/abs/hep-ph/0212150}
  {arXiv:hep-ph/0212150} \BibitemShut {NoStop}%
\bibitem [{\citenamefont {Carena}\ and\ \citenamefont
  {Haber}(2003)}]{Carena:2002es}%
  \BibitemOpen
  \bibfield  {author} {\bibinfo {author} {\bibfnamefont {M.}~\bibnamefont
  {Carena}}\ and\ \bibinfo {author} {\bibfnamefont {H.~E.}\ \bibnamefont
  {Haber}},\ }\href {\doibase 10.1016/S0146-6410(02)00177-1} {\bibfield
  {journal} {\bibinfo  {journal} {Prog. Part. Nucl. Phys.}\ }\textbf {\bibinfo
  {volume} {50}},\ \bibinfo {pages} {63} (\bibinfo {year} {2003})},\ \Eprint
  {http://arxiv.org/abs/hep-ph/0208209} {arXiv:hep-ph/0208209} \BibitemShut
  {NoStop}%
\bibitem [{\citenamefont {Lykken}(2010)}]{Lykken:2010mc}%
  \BibitemOpen
  \bibfield  {author} {\bibinfo {author} {\bibfnamefont {J.~D.}\ \bibnamefont
  {Lykken}}\ }(\bibinfo {year} {2010})\ \Eprint
  {http://arxiv.org/abs/1005.1676} {arXiv:1005.1676 [hep-ph]} \BibitemShut
  {NoStop}%
\bibitem [{\citenamefont {Pati}\ and\ \citenamefont
  {Salam}(1974)}]{Pati:1974yy}%
  \BibitemOpen
  \bibfield  {author} {\bibinfo {author} {\bibfnamefont {J.~C.}\ \bibnamefont
  {Pati}}\ and\ \bibinfo {author} {\bibfnamefont {A.}~\bibnamefont {Salam}},\
  }\href {\doibase 10.1103/PhysRevD.10.275} {\bibfield  {journal} {\bibinfo
  {journal} {Phys. Rev. D}\ }\textbf {\bibinfo {volume} {10}},\ \bibinfo
  {pages} {275} (\bibinfo {year} {1974})},\ \bibinfo {note} {[Erratum:
  Phys.Rev.D 11, 703--703 (1975)]}\BibitemShut {NoStop}%
\bibitem [{\citenamefont {Mohapatra}\ and\ \citenamefont
  {Pati}(1975)}]{Mohapatra:1974gc}%
  \BibitemOpen
  \bibfield  {author} {\bibinfo {author} {\bibfnamefont {R.~N.}\ \bibnamefont
  {Mohapatra}}\ and\ \bibinfo {author} {\bibfnamefont {J.~C.}\ \bibnamefont
  {Pati}},\ }\href {\doibase 10.1103/PhysRevD.11.2558} {\bibfield  {journal}
  {\bibinfo  {journal} {Phys. Rev. D}\ }\textbf {\bibinfo {volume} {11}},\
  \bibinfo {pages} {2558} (\bibinfo {year} {1975})}\BibitemShut {NoStop}%
\bibitem [{\citenamefont {Senjanovic}\ and\ \citenamefont
  {Mohapatra}(1975)}]{Senjanovic:1975rk}%
  \BibitemOpen
  \bibfield  {author} {\bibinfo {author} {\bibfnamefont {G.}~\bibnamefont
  {Senjanovic}}\ and\ \bibinfo {author} {\bibfnamefont {R.~N.}\ \bibnamefont
  {Mohapatra}},\ }\href {\doibase 10.1103/PhysRevD.12.1502} {\bibfield
  {journal} {\bibinfo  {journal} {Phys. Rev. D}\ }\textbf {\bibinfo {volume}
  {12}},\ \bibinfo {pages} {1502} (\bibinfo {year} {1975})}\BibitemShut
  {NoStop}%
\bibitem [{\citenamefont {Mohapatra}\ \emph {et~al.}(1978)\citenamefont
  {Mohapatra}, \citenamefont {Paige},\ and\ \citenamefont
  {Sidhu}}]{Mohapatra:1977mj}%
  \BibitemOpen
  \bibfield  {author} {\bibinfo {author} {\bibfnamefont {R.~N.}\ \bibnamefont
  {Mohapatra}}, \bibinfo {author} {\bibfnamefont {F.~E.}\ \bibnamefont
  {Paige}}, \ and\ \bibinfo {author} {\bibfnamefont {D.~P.}\ \bibnamefont
  {Sidhu}},\ }\href {\doibase 10.1103/PhysRevD.17.2462} {\bibfield  {journal}
  {\bibinfo  {journal} {Phys. Rev. D}\ }\textbf {\bibinfo {volume} {17}},\
  \bibinfo {pages} {2462} (\bibinfo {year} {1978})}\BibitemShut {NoStop}%
\bibitem [{\citenamefont {Bhupal~Dev}\ \emph {et~al.}(2018)\citenamefont
  {Bhupal~Dev}, \citenamefont {Mohapatra},\ and\ \citenamefont
  {Zhang}}]{BhupalDev:2018vpr}%
  \BibitemOpen
  \bibfield  {author} {\bibinfo {author} {\bibfnamefont {P.~S.}\ \bibnamefont
  {Bhupal~Dev}}, \bibinfo {author} {\bibfnamefont {R.~N.}\ \bibnamefont
  {Mohapatra}}, \ and\ \bibinfo {author} {\bibfnamefont {Y.}~\bibnamefont
  {Zhang}},\ }\href {\doibase 10.1103/PhysRevD.98.075028} {\bibfield  {journal}
  {\bibinfo  {journal} {Phys. Rev. D}\ }\textbf {\bibinfo {volume} {98}},\
  \bibinfo {pages} {075028} (\bibinfo {year} {2018})},\ \Eprint
  {http://arxiv.org/abs/1803.11167} {arXiv:1803.11167 [hep-ph]} \BibitemShut
  {NoStop}%
\bibitem [{\citenamefont {Lee}(1973)}]{Lee:1973iz}%
  \BibitemOpen
  \bibfield  {author} {\bibinfo {author} {\bibfnamefont {T.~D.}\ \bibnamefont
  {Lee}},\ }\href {\doibase 10.1103/PhysRevD.8.1226} {\bibfield  {journal}
  {\bibinfo  {journal} {Phys. Rev. D}\ }\textbf {\bibinfo {volume} {8}},\
  \bibinfo {pages} {1226} (\bibinfo {year} {1973})}\BibitemShut {NoStop}%
\bibitem [{\citenamefont {Haber}\ and\ \citenamefont
  {Kane}(1985)}]{Haber:1984rc}%
  \BibitemOpen
  \bibfield  {author} {\bibinfo {author} {\bibfnamefont {H.~E.}\ \bibnamefont
  {Haber}}\ and\ \bibinfo {author} {\bibfnamefont {G.~L.}\ \bibnamefont
  {Kane}},\ }\href {\doibase 10.1016/0370-1573(85)90051-1} {\bibfield
  {journal} {\bibinfo  {journal} {Phys. Rept.}\ }\textbf {\bibinfo {volume}
  {117}},\ \bibinfo {pages} {75} (\bibinfo {year} {1985})}\BibitemShut
  {NoStop}%
\bibitem [{\citenamefont {Gunion}\ and\ \citenamefont
  {Haber}(1986)}]{Gunion:1984yn}%
  \BibitemOpen
  \bibfield  {author} {\bibinfo {author} {\bibfnamefont {J.~F.}\ \bibnamefont
  {Gunion}}\ and\ \bibinfo {author} {\bibfnamefont {H.~E.}\ \bibnamefont
  {Haber}},\ }\href {\doibase 10.1016/0550-3213(86)90340-8} {\bibfield
  {journal} {\bibinfo  {journal} {Nucl. Phys. B}\ }\textbf {\bibinfo {volume}
  {272}},\ \bibinfo {pages} {1} (\bibinfo {year} {1986})},\ \bibinfo {note}
  {[Erratum: Nucl.Phys.B 402, 567--569 (1993)]}\BibitemShut {NoStop}%
\bibitem [{\citenamefont {Branco}\ \emph {et~al.}(2012)\citenamefont {Branco},
  \citenamefont {Ferreira}, \citenamefont {Lavoura}, \citenamefont {Rebelo},
  \citenamefont {Sher},\ and\ \citenamefont {Silva}}]{Branco:2011iw}%
  \BibitemOpen
  \bibfield  {author} {\bibinfo {author} {\bibfnamefont {G.~C.}\ \bibnamefont
  {Branco}}, \bibinfo {author} {\bibfnamefont {P.~M.}\ \bibnamefont
  {Ferreira}}, \bibinfo {author} {\bibfnamefont {L.}~\bibnamefont {Lavoura}},
  \bibinfo {author} {\bibfnamefont {M.~N.}\ \bibnamefont {Rebelo}}, \bibinfo
  {author} {\bibfnamefont {M.}~\bibnamefont {Sher}}, \ and\ \bibinfo {author}
  {\bibfnamefont {J.~P.}\ \bibnamefont {Silva}},\ }\href {\doibase
  10.1016/j.physrep.2012.02.002} {\bibfield  {journal} {\bibinfo  {journal}
  {Phys. Rept.}\ }\textbf {\bibinfo {volume} {516}},\ \bibinfo {pages} {1}
  (\bibinfo {year} {2012})},\ \Eprint {http://arxiv.org/abs/1106.0034}
  {arXiv:1106.0034 [hep-ph]} \BibitemShut {NoStop}%
\bibitem [{\citenamefont {Black}\ \emph {et~al.}(2023)\citenamefont {Black},
  \citenamefont {Plascencia},\ and\ \citenamefont
  {Tetlalmatzi-Xolocotzi}}]{Black:2022wbg}%
  \BibitemOpen
  \bibfield  {author} {\bibinfo {author} {\bibfnamefont {M.}~\bibnamefont
  {Black}}, \bibinfo {author} {\bibfnamefont {A.~D.}\ \bibnamefont
  {Plascencia}}, \ and\ \bibinfo {author} {\bibfnamefont {G.}~\bibnamefont
  {Tetlalmatzi-Xolocotzi}},\ }\href {\doibase 10.1103/PhysRevD.107.035013}
  {\bibfield  {journal} {\bibinfo  {journal} {Phys. Rev. D}\ }\textbf {\bibinfo
  {volume} {107}},\ \bibinfo {pages} {035013} (\bibinfo {year} {2023})},\
  \Eprint {http://arxiv.org/abs/2208.08995} {arXiv:2208.08995 [hep-ph]}
  \BibitemShut {NoStop}%
\bibitem [{\citenamefont {Konetschny}\ and\ \citenamefont
  {Kummer}(1977)}]{Konetschny:1977bn}%
  \BibitemOpen
  \bibfield  {author} {\bibinfo {author} {\bibfnamefont {W.}~\bibnamefont
  {Konetschny}}\ and\ \bibinfo {author} {\bibfnamefont {W.}~\bibnamefont
  {Kummer}},\ }\href {\doibase 10.1016/0370-2693(77)90407-5} {\bibfield
  {journal} {\bibinfo  {journal} {Phys. Lett. B}\ }\textbf {\bibinfo {volume}
  {70}},\ \bibinfo {pages} {433} (\bibinfo {year} {1977})}\BibitemShut
  {NoStop}%
\bibitem [{\citenamefont {Magg}\ and\ \citenamefont
  {Wetterich}(1980)}]{Magg:1980ut}%
  \BibitemOpen
  \bibfield  {author} {\bibinfo {author} {\bibfnamefont {M.}~\bibnamefont
  {Magg}}\ and\ \bibinfo {author} {\bibfnamefont {C.}~\bibnamefont
  {Wetterich}},\ }\href {\doibase 10.1016/0370-2693(80)90825-4} {\bibfield
  {journal} {\bibinfo  {journal} {Phys. Lett. B}\ }\textbf {\bibinfo {volume}
  {94}},\ \bibinfo {pages} {61} (\bibinfo {year} {1980})}\BibitemShut {NoStop}%
\bibitem [{\citenamefont {Schechter}\ and\ \citenamefont
  {Valle}(1980)}]{Schechter:1980gr}%
  \BibitemOpen
  \bibfield  {author} {\bibinfo {author} {\bibfnamefont {J.}~\bibnamefont
  {Schechter}}\ and\ \bibinfo {author} {\bibfnamefont {J.~W.~F.}\ \bibnamefont
  {Valle}},\ }\href {\doibase 10.1103/PhysRevD.22.2227} {\bibfield  {journal}
  {\bibinfo  {journal} {Phys. Rev. D}\ }\textbf {\bibinfo {volume} {22}},\
  \bibinfo {pages} {2227} (\bibinfo {year} {1980})}\BibitemShut {NoStop}%
\bibitem [{\citenamefont {Cheng}\ and\ \citenamefont
  {Li}(1980)}]{Cheng:1980qt}%
  \BibitemOpen
  \bibfield  {author} {\bibinfo {author} {\bibfnamefont {T.~P.}\ \bibnamefont
  {Cheng}}\ and\ \bibinfo {author} {\bibfnamefont {L.-F.}\ \bibnamefont {Li}},\
  }\href {\doibase 10.1103/PhysRevD.22.2860} {\bibfield  {journal} {\bibinfo
  {journal} {Phys. Rev. D}\ }\textbf {\bibinfo {volume} {22}},\ \bibinfo
  {pages} {2860} (\bibinfo {year} {1980})}\BibitemShut {NoStop}%
\bibitem [{\citenamefont {Lazarides}\ \emph {et~al.}(1981)\citenamefont
  {Lazarides}, \citenamefont {Shafi},\ and\ \citenamefont
  {Wetterich}}]{Lazarides:1980nt}%
  \BibitemOpen
  \bibfield  {author} {\bibinfo {author} {\bibfnamefont {G.}~\bibnamefont
  {Lazarides}}, \bibinfo {author} {\bibfnamefont {Q.}~\bibnamefont {Shafi}}, \
  and\ \bibinfo {author} {\bibfnamefont {C.}~\bibnamefont {Wetterich}},\ }\href
  {\doibase 10.1016/0550-3213(81)90354-0} {\bibfield  {journal} {\bibinfo
  {journal} {Nucl. Phys. B}\ }\textbf {\bibinfo {volume} {181}},\ \bibinfo
  {pages} {287} (\bibinfo {year} {1981})}\BibitemShut {NoStop}%
\bibitem [{\citenamefont {Baer}\ \emph {et~al.}(2013)\citenamefont {Baer} \emph
  {et~al.}}]{ILC:2013jhg}%
  \BibitemOpen
  \bibfield  {author} {\bibinfo {author} {\bibfnamefont {H.}~\bibnamefont
  {Baer}} \emph {et~al.} (\bibinfo {collaboration} {ILC}),\ }\href@noop {} {\
  (\bibinfo {year} {2013})},\ \Eprint {http://arxiv.org/abs/1306.6352}
  {arXiv:1306.6352 [hep-ph]} \BibitemShut {NoStop}%
\bibitem [{\citenamefont {Evans}\ and\ \citenamefont
  {Michizono}(2017)}]{Evans:2017rvt}%
  \BibitemOpen
  \bibfield  {author} {\bibinfo {author} {\bibfnamefont {L.}~\bibnamefont
  {Evans}}\ and\ \bibinfo {author} {\bibfnamefont {S.}~\bibnamefont
  {Michizono}} (\bibinfo {collaboration} {Linear Collide}),\ }\href@noop {} {\
  (\bibinfo {year} {2017})},\ \Eprint {http://arxiv.org/abs/1711.00568}
  {arXiv:1711.00568 [physics.acc-ph]} \BibitemShut {NoStop}%
\bibitem [{\citenamefont {Accomando}\ \emph {et~al.}(2004)\citenamefont
  {Accomando} \emph {et~al.}}]{CLICPhysicsWorkingGroup:2004qvu}%
  \BibitemOpen
  \bibfield  {author} {\bibinfo {author} {\bibfnamefont {E.}~\bibnamefont
  {Accomando}} \emph {et~al.} (\bibinfo {collaboration} {CLIC Physics Working
  Group}),\ }in\ \href {\doibase 10.5170/CERN-2004-005} {\emph {\bibinfo
  {booktitle} {{11th International Conference on Hadron Spectroscopy}}}},\
  \bibinfo {series and number} {CERN Yellow Reports: Monographs}\ (\bibinfo
  {year} {2004})\ \Eprint {http://arxiv.org/abs/hep-ph/0412251}
  {arXiv:hep-ph/0412251} \BibitemShut {NoStop}%
\bibitem [{\citenamefont {Boland}\ \emph {et~al.}(2016)\citenamefont {Boland}
  \emph {et~al.}}]{CLIC:2016zwp}%
  \BibitemOpen
  \bibfield  {author} {\bibinfo {author} {\bibfnamefont {M.~J.}\ \bibnamefont
  {Boland}} \emph {et~al.} (\bibinfo {collaboration} {CLIC, CLICdp}),\ }\href
  {\doibase 10.5170/CERN-2016-004} {\  (\bibinfo {year} {2016}),\
  10.5170/CERN-2016-004},\ \Eprint {http://arxiv.org/abs/1608.07537}
  {arXiv:1608.07537 [physics.acc-ph]} \BibitemShut {NoStop}%
\bibitem [{\citenamefont {Bellgardt}\ \emph {et~al.}(1988)\citenamefont
  {Bellgardt} \emph {et~al.}}]{SINDRUM:1987nra}%
  \BibitemOpen
  \bibfield  {author} {\bibinfo {author} {\bibfnamefont {U.}~\bibnamefont
  {Bellgardt}} \emph {et~al.} (\bibinfo {collaboration} {SINDRUM}),\ }\href
  {\doibase 10.1016/0550-3213(88)90462-2} {\bibfield  {journal} {\bibinfo
  {journal} {Nucl. Phys. B}\ }\textbf {\bibinfo {volume} {299}},\ \bibinfo
  {pages} {1} (\bibinfo {year} {1988})}\BibitemShut {NoStop}%
\bibitem [{\citenamefont {Baldini}\ \emph {et~al.}(2016)\citenamefont {Baldini}
  \emph {et~al.}}]{MEG:2016leq}%
  \BibitemOpen
  \bibfield  {author} {\bibinfo {author} {\bibfnamefont {A.~M.}\ \bibnamefont
  {Baldini}} \emph {et~al.} (\bibinfo {collaboration} {MEG}),\ }\href {\doibase
  10.1140/epjc/s10052-016-4271-x} {\bibfield  {journal} {\bibinfo  {journal}
  {Eur. Phys. J. C}\ }\textbf {\bibinfo {volume} {76}},\ \bibinfo {pages} {434}
  (\bibinfo {year} {2016})},\ \Eprint {http://arxiv.org/abs/1605.05081}
  {arXiv:1605.05081 [hep-ex]} \BibitemShut {NoStop}%
\bibitem [{\citenamefont {Workman}\ \emph {et~al.}(2022)\citenamefont {Workman}
  \emph {et~al.}}]{Workman:2022ynf}%
  \BibitemOpen
  \bibfield  {author} {\bibinfo {author} {\bibfnamefont {R.~L.}\ \bibnamefont
  {Workman}} \emph {et~al.} (\bibinfo {collaboration} {Particle Data Group}),\
  }\href {\doibase 10.1093/ptep/ptac097} {\bibfield  {journal} {\bibinfo
  {journal} {PTEP}\ }\textbf {\bibinfo {volume} {2022}},\ \bibinfo {pages}
  {083C01} (\bibinfo {year} {2022})}\BibitemShut {NoStop}%
\bibitem [{\citenamefont {Amhis}\ \emph {et~al.}(2017)\citenamefont {Amhis}
  \emph {et~al.}}]{HFLAV:2016hnz}%
  \BibitemOpen
  \bibfield  {author} {\bibinfo {author} {\bibfnamefont {Y.}~\bibnamefont
  {Amhis}} \emph {et~al.} (\bibinfo {collaboration} {HFLAV}),\ }\href {\doibase
  10.1140/epjc/s10052-017-5058-4} {\bibfield  {journal} {\bibinfo  {journal}
  {Eur. Phys. J. C}\ }\textbf {\bibinfo {volume} {77}},\ \bibinfo {pages} {895}
  (\bibinfo {year} {2017})},\ \Eprint {http://arxiv.org/abs/1612.07233}
  {arXiv:1612.07233 [hep-ex]} \BibitemShut {NoStop}%
\bibitem [{\citenamefont {Willmann}\ \emph {et~al.}(1999)\citenamefont
  {Willmann} \emph {et~al.}}]{Willmann:1998gd}%
  \BibitemOpen
  \bibfield  {author} {\bibinfo {author} {\bibfnamefont {L.}~\bibnamefont
  {Willmann}} \emph {et~al.},\ }\href {\doibase 10.1103/PhysRevLett.82.49}
  {\bibfield  {journal} {\bibinfo  {journal} {Phys. Rev. Lett.}\ }\textbf
  {\bibinfo {volume} {82}},\ \bibinfo {pages} {49} (\bibinfo {year} {1999})},\
  \Eprint {http://arxiv.org/abs/hep-ex/9807011} {arXiv:hep-ex/9807011}
  \BibitemShut {NoStop}%
\bibitem [{\citenamefont {Abdallah}\ \emph {et~al.}(2006)\citenamefont
  {Abdallah} \emph {et~al.}}]{DELPHI:2005wxt}%
  \BibitemOpen
  \bibfield  {author} {\bibinfo {author} {\bibfnamefont {J.}~\bibnamefont
  {Abdallah}} \emph {et~al.} (\bibinfo {collaboration} {DELPHI}),\ }\href
  {\doibase 10.1140/epjc/s2005-02461-0} {\bibfield  {journal} {\bibinfo
  {journal} {Eur. Phys. J. C}\ }\textbf {\bibinfo {volume} {45}},\ \bibinfo
  {pages} {589} (\bibinfo {year} {2006})},\ \Eprint
  {http://arxiv.org/abs/hep-ex/0512012} {arXiv:hep-ex/0512012} \BibitemShut
  {NoStop}%
\bibitem [{ATL(2021)}]{ATLAS-CONF-2021-011}%
  \BibitemOpen
  \href {http://cds.cern.ch/record/2759285} {\emph {\bibinfo {title} {{Search
  for new phenomena in three- or four-lepton events in $pp$ collisions at
  $\sqrt{s} = $ 13 TeV with the ATLAS detector}}}},\ \bibinfo {type} {Tech.
  Rep.}\ (\bibinfo  {institution} {CERN},\ \bibinfo {address} {Geneva},\
  \bibinfo {year} {2021})\BibitemShut {NoStop}%
\bibitem [{ATL(2022)}]{ATLAS:2022pbd}%
  \BibitemOpen
  \href@noop {} {\bibfield  {journal} {\bibinfo  {journal} {CERN-EP-2022-212}\
  } (\bibinfo {year} {2022})},\ \Eprint {http://arxiv.org/abs/2211.07505}
  {arXiv:2211.07505 [hep-ex]} \BibitemShut {NoStop}%
\bibitem [{\citenamefont {Parker}\ \emph {et~al.}(2018)\citenamefont {Parker},
  \citenamefont {Yu}, \citenamefont {Zhong}, \citenamefont {Estey},\ and\
  \citenamefont {M{\"u}ller}}]{Parker2018MeasurementOT}%
  \BibitemOpen
  \bibfield  {author} {\bibinfo {author} {\bibfnamefont {R.~H.}\ \bibnamefont
  {Parker}}, \bibinfo {author} {\bibfnamefont {C.}~\bibnamefont {Yu}}, \bibinfo
  {author} {\bibfnamefont {W.}~\bibnamefont {Zhong}}, \bibinfo {author}
  {\bibfnamefont {B.}~\bibnamefont {Estey}}, \ and\ \bibinfo {author}
  {\bibfnamefont {H.}~\bibnamefont {M{\"u}ller}},\ }\href@noop {} {\bibfield
  {journal} {\bibinfo  {journal} {Science}\ }\textbf {\bibinfo {volume}
  {360}},\ \bibinfo {pages} {191 } (\bibinfo {year} {2018})}\BibitemShut
  {NoStop}%
\bibitem [{\citenamefont {Morel}\ \emph {et~al.}(2020)\citenamefont {Morel},
  \citenamefont {Yao}, \citenamefont {Clad{\'e}},\ and\ \citenamefont
  {Guellati-Kh{\'e}lifa}}]{morel2020determination}%
  \BibitemOpen
  \bibfield  {author} {\bibinfo {author} {\bibfnamefont {L.}~\bibnamefont
  {Morel}}, \bibinfo {author} {\bibfnamefont {Z.}~\bibnamefont {Yao}}, \bibinfo
  {author} {\bibfnamefont {P.}~\bibnamefont {Clad{\'e}}}, \ and\ \bibinfo
  {author} {\bibfnamefont {S.}~\bibnamefont {Guellati-Kh{\'e}lifa}},\
  }\href@noop {} {\bibfield  {journal} {\bibinfo  {journal} {Nature}\ }\textbf
  {\bibinfo {volume} {588}},\ \bibinfo {pages} {61} (\bibinfo {year}
  {2020})}\BibitemShut {NoStop}%
\bibitem [{\citenamefont {Abi}\ \emph {et~al.}(2021)\citenamefont {Abi} \emph
  {et~al.}}]{Muong-2:2021ojo}%
  \BibitemOpen
  \bibfield  {author} {\bibinfo {author} {\bibfnamefont {B.}~\bibnamefont
  {Abi}} \emph {et~al.} (\bibinfo {collaboration} {Muon g-2}),\ }\href
  {\doibase 10.1103/PhysRevLett.126.141801} {\bibfield  {journal} {\bibinfo
  {journal} {Phys. Rev. Lett.}\ }\textbf {\bibinfo {volume} {126}},\ \bibinfo
  {pages} {141801} (\bibinfo {year} {2021})},\ \Eprint
  {http://arxiv.org/abs/2104.03281} {arXiv:2104.03281 [hep-ex]} \BibitemShut
  {NoStop}%
\bibitem [{\citenamefont {Queiroz}\ and\ \citenamefont
  {Shepherd}(2014)}]{Queiroz:2014zfa}%
  \BibitemOpen
  \bibfield  {author} {\bibinfo {author} {\bibfnamefont {F.~S.}\ \bibnamefont
  {Queiroz}}\ and\ \bibinfo {author} {\bibfnamefont {W.}~\bibnamefont
  {Shepherd}},\ }\href {\doibase 10.1103/PhysRevD.89.095024} {\bibfield
  {journal} {\bibinfo  {journal} {Phys. Rev. D}\ }\textbf {\bibinfo {volume}
  {89}},\ \bibinfo {pages} {095024} (\bibinfo {year} {2014})},\ \Eprint
  {http://arxiv.org/abs/1403.2309} {arXiv:1403.2309 [hep-ph]} \BibitemShut
  {NoStop}%
\bibitem [{\citenamefont {Lindner}\ \emph {et~al.}(2018)\citenamefont
  {Lindner}, \citenamefont {Platscher},\ and\ \citenamefont
  {Queiroz}}]{Lindner:2016bgg}%
  \BibitemOpen
  \bibfield  {author} {\bibinfo {author} {\bibfnamefont {M.}~\bibnamefont
  {Lindner}}, \bibinfo {author} {\bibfnamefont {M.}~\bibnamefont {Platscher}},
  \ and\ \bibinfo {author} {\bibfnamefont {F.~S.}\ \bibnamefont {Queiroz}},\
  }\href {\doibase 10.1016/j.physrep.2017.12.001} {\bibfield  {journal}
  {\bibinfo  {journal} {Phys. Rept.}\ }\textbf {\bibinfo {volume} {731}},\
  \bibinfo {pages} {1} (\bibinfo {year} {2018})},\ \Eprint
  {http://arxiv.org/abs/1610.06587} {arXiv:1610.06587 [hep-ph]} \BibitemShut
  {NoStop}%
\bibitem [{\citenamefont {Yu}\ \emph {et~al.}(2017)\citenamefont {Yu},
  \citenamefont {Ruan}, \citenamefont {Boudry},\ and\ \citenamefont
  {Videau}}]{Yu:2017mpx}%
  \BibitemOpen
  \bibfield  {author} {\bibinfo {author} {\bibfnamefont {D.}~\bibnamefont
  {Yu}}, \bibinfo {author} {\bibfnamefont {M.}~\bibnamefont {Ruan}}, \bibinfo
  {author} {\bibfnamefont {V.}~\bibnamefont {Boudry}}, \ and\ \bibinfo {author}
  {\bibfnamefont {H.}~\bibnamefont {Videau}},\ }\href {\doibase
  10.1140/epjc/s10052-017-5146-5} {\bibfield  {journal} {\bibinfo  {journal}
  {Eur. Phys. J. C}\ }\textbf {\bibinfo {volume} {77}},\ \bibinfo {pages} {591}
  (\bibinfo {year} {2017})},\ \Eprint {http://arxiv.org/abs/1701.07542}
  {arXiv:1701.07542 [physics.ins-det]} \BibitemShut {NoStop}%
\bibitem [{\citenamefont {Alwall}\ \emph {et~al.}(2014)\citenamefont {Alwall},
  \citenamefont {Frederix}, \citenamefont {Frixione}, \citenamefont {Hirschi},
  \citenamefont {Maltoni}, \citenamefont {Mattelaer}, \citenamefont {Shao},
  \citenamefont {Stelzer}, \citenamefont {Torrielli},\ and\ \citenamefont
  {Zaro}}]{Alwall:2014hca}%
  \BibitemOpen
  \bibfield  {author} {\bibinfo {author} {\bibfnamefont {J.}~\bibnamefont
  {Alwall}}, \bibinfo {author} {\bibfnamefont {R.}~\bibnamefont {Frederix}},
  \bibinfo {author} {\bibfnamefont {S.}~\bibnamefont {Frixione}}, \bibinfo
  {author} {\bibfnamefont {V.}~\bibnamefont {Hirschi}}, \bibinfo {author}
  {\bibfnamefont {F.}~\bibnamefont {Maltoni}}, \bibinfo {author} {\bibfnamefont
  {O.}~\bibnamefont {Mattelaer}}, \bibinfo {author} {\bibfnamefont {H.~S.}\
  \bibnamefont {Shao}}, \bibinfo {author} {\bibfnamefont {T.}~\bibnamefont
  {Stelzer}}, \bibinfo {author} {\bibfnamefont {P.}~\bibnamefont {Torrielli}},
  \ and\ \bibinfo {author} {\bibfnamefont {M.}~\bibnamefont {Zaro}},\ }\href
  {\doibase 10.1007/JHEP07(2014)079} {\bibfield  {journal} {\bibinfo  {journal}
  {JHEP}\ }\textbf {\bibinfo {volume} {07}},\ \bibinfo {pages} {079} (\bibinfo
  {year} {2014})},\ \Eprint {http://arxiv.org/abs/1405.0301} {arXiv:1405.0301
  [hep-ph]} \BibitemShut {NoStop}%
\bibitem [{\citenamefont {Cai}\ \emph {et~al.}(2018)\citenamefont {Cai},
  \citenamefont {Han}, \citenamefont {Li},\ and\ \citenamefont
  {Ruiz}}]{Cai:2017mow}%
  \BibitemOpen
  \bibfield  {author} {\bibinfo {author} {\bibfnamefont {Y.}~\bibnamefont
  {Cai}}, \bibinfo {author} {\bibfnamefont {T.}~\bibnamefont {Han}}, \bibinfo
  {author} {\bibfnamefont {T.}~\bibnamefont {Li}}, \ and\ \bibinfo {author}
  {\bibfnamefont {R.}~\bibnamefont {Ruiz}},\ }\href {\doibase
  10.3389/fphy.2018.00040} {\bibfield  {journal} {\bibinfo  {journal} {Front.
  in Phys.}\ }\textbf {\bibinfo {volume} {6}},\ \bibinfo {pages} {40} (\bibinfo
  {year} {2018})},\ \Eprint {http://arxiv.org/abs/1711.02180} {arXiv:1711.02180
  [hep-ph]} \BibitemShut {NoStop}%
\bibitem [{\citenamefont {Bhupal~Dev}\ and\ \citenamefont
  {Zhang}(2018)}]{BhupalDev:2018tox}%
  \BibitemOpen
  \bibfield  {author} {\bibinfo {author} {\bibfnamefont {P.~S.}\ \bibnamefont
  {Bhupal~Dev}}\ and\ \bibinfo {author} {\bibfnamefont {Y.}~\bibnamefont
  {Zhang}},\ }\href {\doibase 10.1007/JHEP10(2018)199} {\bibfield  {journal}
  {\bibinfo  {journal} {JHEP}\ }\textbf {\bibinfo {volume} {10}},\ \bibinfo
  {pages} {199} (\bibinfo {year} {2018})},\ \Eprint
  {http://arxiv.org/abs/1808.00943} {arXiv:1808.00943 [hep-ph]} \BibitemShut
  {NoStop}%
\bibitem [{\citenamefont {Akeroyd}\ and\ \citenamefont
  {Aoki}(2005)}]{Akeroyd:2005gt}%
  \BibitemOpen
  \bibfield  {author} {\bibinfo {author} {\bibfnamefont {A.~G.}\ \bibnamefont
  {Akeroyd}}\ and\ \bibinfo {author} {\bibfnamefont {M.}~\bibnamefont {Aoki}},\
  }\href {\doibase 10.1103/PhysRevD.72.035011} {\bibfield  {journal} {\bibinfo
  {journal} {Phys. Rev. D}\ }\textbf {\bibinfo {volume} {72}},\ \bibinfo
  {pages} {035011} (\bibinfo {year} {2005})},\ \Eprint
  {http://arxiv.org/abs/hep-ph/0506176} {arXiv:hep-ph/0506176} \BibitemShut
  {NoStop}%
\bibitem [{\citenamefont {Hektor}\ \emph {et~al.}(2007)\citenamefont {Hektor},
  \citenamefont {Kadastik}, \citenamefont {Muntel}, \citenamefont {Raidal},\
  and\ \citenamefont {Rebane}}]{Hektor:2007uu}%
  \BibitemOpen
  \bibfield  {author} {\bibinfo {author} {\bibfnamefont {A.}~\bibnamefont
  {Hektor}}, \bibinfo {author} {\bibfnamefont {M.}~\bibnamefont {Kadastik}},
  \bibinfo {author} {\bibfnamefont {M.}~\bibnamefont {Muntel}}, \bibinfo
  {author} {\bibfnamefont {M.}~\bibnamefont {Raidal}}, \ and\ \bibinfo {author}
  {\bibfnamefont {L.}~\bibnamefont {Rebane}},\ }\href {\doibase
  10.1016/j.nuclphysb.2007.07.014} {\bibfield  {journal} {\bibinfo  {journal}
  {Nucl. Phys. B}\ }\textbf {\bibinfo {volume} {787}},\ \bibinfo {pages} {198}
  (\bibinfo {year} {2007})},\ \Eprint {http://arxiv.org/abs/0705.1495}
  {arXiv:0705.1495 [hep-ph]} \BibitemShut {NoStop}%
\bibitem [{\citenamefont {Chao}\ \emph {et~al.}(2008)\citenamefont {Chao},
  \citenamefont {Si}, \citenamefont {Xing},\ and\ \citenamefont
  {Zhou}}]{Chao:2008mq}%
  \BibitemOpen
  \bibfield  {author} {\bibinfo {author} {\bibfnamefont {W.}~\bibnamefont
  {Chao}}, \bibinfo {author} {\bibfnamefont {Z.-G.}\ \bibnamefont {Si}},
  \bibinfo {author} {\bibfnamefont {Z.-z.}\ \bibnamefont {Xing}}, \ and\
  \bibinfo {author} {\bibfnamefont {S.}~\bibnamefont {Zhou}},\ }\href {\doibase
  10.1016/j.physletb.2008.08.003} {\bibfield  {journal} {\bibinfo  {journal}
  {Phys. Lett. B}\ }\textbf {\bibinfo {volume} {666}},\ \bibinfo {pages} {451}
  (\bibinfo {year} {2008})},\ \Eprint {http://arxiv.org/abs/0804.1265}
  {arXiv:0804.1265 [hep-ph]} \BibitemShut {NoStop}%
\bibitem [{\citenamefont {Zee}(1985)}]{Zee:1985rj}%
  \BibitemOpen
  \bibfield  {author} {\bibinfo {author} {\bibfnamefont {A.}~\bibnamefont
  {Zee}},\ }\href {\doibase 10.1016/0370-2693(85)90625-2} {\bibfield  {journal}
  {\bibinfo  {journal} {Phys. Lett. B}\ }\textbf {\bibinfo {volume} {161}},\
  \bibinfo {pages} {141} (\bibinfo {year} {1985})}\BibitemShut {NoStop}%
\bibitem [{\citenamefont {Zee}(1986)}]{Zee:1985id}%
  \BibitemOpen
  \bibfield  {author} {\bibinfo {author} {\bibfnamefont {A.}~\bibnamefont
  {Zee}},\ }\href {\doibase 10.1016/0550-3213(86)90475-X} {\bibfield  {journal}
  {\bibinfo  {journal} {Nucl. Phys. B}\ }\textbf {\bibinfo {volume} {264}},\
  \bibinfo {pages} {99} (\bibinfo {year} {1986})}\BibitemShut {NoStop}%
\bibitem [{\citenamefont {Babu}(1988)}]{Babu:1988ki}%
  \BibitemOpen
  \bibfield  {author} {\bibinfo {author} {\bibfnamefont {K.~S.}\ \bibnamefont
  {Babu}},\ }\href {\doibase 10.1016/0370-2693(88)91584-5} {\bibfield
  {journal} {\bibinfo  {journal} {Phys. Lett. B}\ }\textbf {\bibinfo {volume}
  {203}},\ \bibinfo {pages} {132} (\bibinfo {year} {1988})}\BibitemShut
  {NoStop}%
\bibitem [{\citenamefont {Fuks}\ \emph {et~al.}(2020)\citenamefont {Fuks},
  \citenamefont {Nemev\v{s}ek},\ and\ \citenamefont {Ruiz}}]{Fuks:2019clu}%
  \BibitemOpen
  \bibfield  {author} {\bibinfo {author} {\bibfnamefont {B.}~\bibnamefont
  {Fuks}}, \bibinfo {author} {\bibfnamefont {M.}~\bibnamefont {Nemev\v{s}ek}},
  \ and\ \bibinfo {author} {\bibfnamefont {R.}~\bibnamefont {Ruiz}},\ }\href
  {\doibase 10.1103/PhysRevD.101.075022} {\bibfield  {journal} {\bibinfo
  {journal} {Phys. Rev. D}\ }\textbf {\bibinfo {volume} {101}},\ \bibinfo
  {pages} {075022} (\bibinfo {year} {2020})},\ \Eprint
  {http://arxiv.org/abs/1912.08975} {arXiv:1912.08975 [hep-ph]} \BibitemShut
  {NoStop}%
\bibitem [{\citenamefont {Dev}\ \emph {et~al.}(2017)\citenamefont {Dev},
  \citenamefont {Mohapatra},\ and\ \citenamefont {Zhang}}]{Dev:2017dui}%
  \BibitemOpen
  \bibfield  {author} {\bibinfo {author} {\bibfnamefont {P.~S.~B.}\
  \bibnamefont {Dev}}, \bibinfo {author} {\bibfnamefont {R.~N.}\ \bibnamefont
  {Mohapatra}}, \ and\ \bibinfo {author} {\bibfnamefont {Y.}~\bibnamefont
  {Zhang}},\ }\href {\doibase 10.1016/j.nuclphysb.2017.07.021} {\bibfield
  {journal} {\bibinfo  {journal} {Nucl. Phys. B}\ }\textbf {\bibinfo {volume}
  {923}},\ \bibinfo {pages} {179} (\bibinfo {year} {2017})},\ \Eprint
  {http://arxiv.org/abs/1703.02471} {arXiv:1703.02471 [hep-ph]} \BibitemShut
  {NoStop}%
\bibitem [{\citenamefont {Ibe}\ \emph {et~al.}(2022)\citenamefont {Ibe},
  \citenamefont {Kobayashi}, \citenamefont {Nakayama},\ and\ \citenamefont
  {Shirai}}]{Ibe:2021fed}%
  \BibitemOpen
  \bibfield  {author} {\bibinfo {author} {\bibfnamefont {M.}~\bibnamefont
  {Ibe}}, \bibinfo {author} {\bibfnamefont {S.}~\bibnamefont {Kobayashi}},
  \bibinfo {author} {\bibfnamefont {Y.}~\bibnamefont {Nakayama}}, \ and\
  \bibinfo {author} {\bibfnamefont {S.}~\bibnamefont {Shirai}},\ }\href
  {\doibase 10.1007/JHEP03(2022)198} {\bibfield  {journal} {\bibinfo  {journal}
  {JHEP}\ }\textbf {\bibinfo {volume} {03}},\ \bibinfo {pages} {198} (\bibinfo
  {year} {2022})},\ \Eprint {http://arxiv.org/abs/2112.11096} {arXiv:2112.11096
  [hep-ph]} \BibitemShut {NoStop}%
\bibitem [{\citenamefont {Balaji}\ \emph {et~al.}(2022)\citenamefont {Balaji},
  \citenamefont {Dev}, \citenamefont {Silk},\ and\ \citenamefont
  {Zhang}}]{Balaji:2022noj}%
  \BibitemOpen
  \bibfield  {author} {\bibinfo {author} {\bibfnamefont {S.}~\bibnamefont
  {Balaji}}, \bibinfo {author} {\bibfnamefont {P.~S.~B.}\ \bibnamefont {Dev}},
  \bibinfo {author} {\bibfnamefont {J.}~\bibnamefont {Silk}}, \ and\ \bibinfo
  {author} {\bibfnamefont {Y.}~\bibnamefont {Zhang}},\ }\href {\doibase
  10.1088/1475-7516/2022/12/024} {\bibfield  {journal} {\bibinfo  {journal}
  {JCAP}\ }\textbf {\bibinfo {volume} {12}},\ \bibinfo {pages} {024} (\bibinfo
  {year} {2022})},\ \Eprint {http://arxiv.org/abs/2205.01669} {arXiv:2205.01669
  [hep-ph]} \BibitemShut {NoStop}%
\bibitem [{\citenamefont {Feng}\ \emph {et~al.}(2022)\citenamefont {Feng} \emph
  {et~al.}}]{Feng:2022inv}%
  \BibitemOpen
  \bibfield  {author} {\bibinfo {author} {\bibfnamefont {J.~L.}\ \bibnamefont
  {Feng}} \emph {et~al.},\ }\href@noop {} {\  (\bibinfo {year} {2022})},\
  \Eprint {http://arxiv.org/abs/2203.05090} {arXiv:2203.05090 [hep-ex]}
  \BibitemShut {NoStop}%
\bibitem [{\citenamefont {Crivellin}\ \emph {et~al.}(2019)\citenamefont
  {Crivellin}, \citenamefont {Ghezzi}, \citenamefont {Panizzi}, \citenamefont
  {Pruna},\ and\ \citenamefont {Signer}}]{Crivellin:2018ahj}%
  \BibitemOpen
  \bibfield  {author} {\bibinfo {author} {\bibfnamefont {A.}~\bibnamefont
  {Crivellin}}, \bibinfo {author} {\bibfnamefont {M.}~\bibnamefont {Ghezzi}},
  \bibinfo {author} {\bibfnamefont {L.}~\bibnamefont {Panizzi}}, \bibinfo
  {author} {\bibfnamefont {G.~M.}\ \bibnamefont {Pruna}}, \ and\ \bibinfo
  {author} {\bibfnamefont {A.}~\bibnamefont {Signer}},\ }\href {\doibase
  10.1103/PhysRevD.99.035004} {\bibfield  {journal} {\bibinfo  {journal} {Phys.
  Rev. D}\ }\textbf {\bibinfo {volume} {99}},\ \bibinfo {pages} {035004}
  (\bibinfo {year} {2019})},\ \Eprint {http://arxiv.org/abs/1807.10224}
  {arXiv:1807.10224 [hep-ph]} \BibitemShut {NoStop}%
\bibitem [{\citenamefont {Gluza}\ \emph {et~al.}(2021)\citenamefont {Gluza},
  \citenamefont {Kordiaczynska},\ and\ \citenamefont
  {Srivastava}}]{Gluza:2020qrt}%
  \BibitemOpen
  \bibfield  {author} {\bibinfo {author} {\bibfnamefont {J.}~\bibnamefont
  {Gluza}}, \bibinfo {author} {\bibfnamefont {M.}~\bibnamefont
  {Kordiaczynska}}, \ and\ \bibinfo {author} {\bibfnamefont {T.}~\bibnamefont
  {Srivastava}},\ }\href {\doibase 10.1088/1674-1137/abfe51} {\bibfield
  {journal} {\bibinfo  {journal} {Chin. Phys. C}\ }\textbf {\bibinfo {volume}
  {45}},\ \bibinfo {pages} {073113} (\bibinfo {year} {2021})},\ \Eprint
  {http://arxiv.org/abs/2006.04610} {arXiv:2006.04610 [hep-ph]} \BibitemShut
  {NoStop}%
\bibitem [{\citenamefont {Bambhaniya}\ \emph {et~al.}(2015)\citenamefont
  {Bambhaniya}, \citenamefont {Chakrabortty}, \citenamefont {Gluza},
  \citenamefont {Jelinski},\ and\ \citenamefont
  {Szafron}}]{Bambhaniya:2015wna}%
  \BibitemOpen
  \bibfield  {author} {\bibinfo {author} {\bibfnamefont {G.}~\bibnamefont
  {Bambhaniya}}, \bibinfo {author} {\bibfnamefont {J.}~\bibnamefont
  {Chakrabortty}}, \bibinfo {author} {\bibfnamefont {J.}~\bibnamefont {Gluza}},
  \bibinfo {author} {\bibfnamefont {T.}~\bibnamefont {Jelinski}}, \ and\
  \bibinfo {author} {\bibfnamefont {R.}~\bibnamefont {Szafron}},\ }\href
  {\doibase 10.1103/PhysRevD.92.015016} {\bibfield  {journal} {\bibinfo
  {journal} {Phys. Rev. D}\ }\textbf {\bibinfo {volume} {92}},\ \bibinfo
  {pages} {015016} (\bibinfo {year} {2015})},\ \Eprint
  {http://arxiv.org/abs/1504.03999} {arXiv:1504.03999 [hep-ph]} \BibitemShut
  {NoStop}%
\bibitem [{CMS(2022)}]{CMS-PAS-FTR-22-006}%
  \BibitemOpen
  \href {https://cds.cern.ch/record/2808604} {\emph {\bibinfo {title}
  {{Prospects for a Search for Doubly Charged Higgs Bosons at the HL-LHC}}}},\
  \bibinfo {type} {Tech. Rep.}\ (\bibinfo  {institution} {CERN},\ \bibinfo
  {address} {Geneva},\ \bibinfo {year} {2022})\BibitemShut {NoStop}%
\bibitem [{\citenamefont {Ruiz}(2022)}]{Ruiz:2022sct}%
  \BibitemOpen
  \bibfield  {author} {\bibinfo {author} {\bibfnamefont {R.}~\bibnamefont
  {Ruiz}},\ }\href {\doibase 10.1007/JHEP10(2022)200} {\bibfield  {journal}
  {\bibinfo  {journal} {JHEP}\ }\textbf {\bibinfo {volume} {10}},\ \bibinfo
  {pages} {200} (\bibinfo {year} {2022})},\ \Eprint
  {http://arxiv.org/abs/2206.14833} {arXiv:2206.14833 [hep-ph]} \BibitemShut
  {NoStop}%
\bibitem [{\citenamefont {Abada}\ \emph {et~al.}(2019)\citenamefont {Abada}
  \emph {et~al.}}]{FCC:2018vvp}%
  \BibitemOpen
  \bibfield  {author} {\bibinfo {author} {\bibfnamefont {A.}~\bibnamefont
  {Abada}} \emph {et~al.} (\bibinfo {collaboration} {FCC}),\ }\href {\doibase
  10.1140/epjst/e2019-900087-0} {\bibfield  {journal} {\bibinfo  {journal}
  {Eur. Phys. J. ST}\ }\textbf {\bibinfo {volume} {228}},\ \bibinfo {pages}
  {755} (\bibinfo {year} {2019})}\BibitemShut {NoStop}%
\bibitem [{\citenamefont {Dev}\ \emph {et~al.}(2018)\citenamefont {Dev},
  \citenamefont {Ramsey-Musolf},\ and\ \citenamefont {Zhang}}]{Dev:2018sel}%
  \BibitemOpen
  \bibfield  {author} {\bibinfo {author} {\bibfnamefont {P.~S.~B.}\
  \bibnamefont {Dev}}, \bibinfo {author} {\bibfnamefont {M.~J.}\ \bibnamefont
  {Ramsey-Musolf}}, \ and\ \bibinfo {author} {\bibfnamefont {Y.}~\bibnamefont
  {Zhang}},\ }\href {\doibase 10.1103/PhysRevD.98.055013} {\bibfield  {journal}
  {\bibinfo  {journal} {Phys. Rev. D}\ }\textbf {\bibinfo {volume} {98}},\
  \bibinfo {pages} {055013} (\bibinfo {year} {2018})},\ \Eprint
  {http://arxiv.org/abs/1806.08499} {arXiv:1806.08499 [hep-ph]} \BibitemShut
  {NoStop}%
\bibitem [{\citenamefont {Aoyama}\ \emph {et~al.}(2020)\citenamefont {Aoyama}
  \emph {et~al.}}]{Aoyama:2020ynm}%
  \BibitemOpen
  \bibfield  {author} {\bibinfo {author} {\bibfnamefont {T.}~\bibnamefont
  {Aoyama}} \emph {et~al.},\ }\href {\doibase 10.1016/j.physrep.2020.07.006}
  {\bibfield  {journal} {\bibinfo  {journal} {Phys. Rept.}\ }\textbf {\bibinfo
  {volume} {887}},\ \bibinfo {pages} {1} (\bibinfo {year} {2020})},\ \Eprint
  {http://arxiv.org/abs/2006.04822} {arXiv:2006.04822 [hep-ph]} \BibitemShut
  {NoStop}%
\bibitem [{\citenamefont {Borsanyi}\ \emph {et~al.}(2021)\citenamefont
  {Borsanyi} \emph {et~al.}}]{Borsanyi:2020mff}%
  \BibitemOpen
  \bibfield  {author} {\bibinfo {author} {\bibfnamefont {S.}~\bibnamefont
  {Borsanyi}} \emph {et~al.},\ }\href {\doibase 10.1038/s41586-021-03418-1}
  {\bibfield  {journal} {\bibinfo  {journal} {Nature}\ }\textbf {\bibinfo
  {volume} {593}},\ \bibinfo {pages} {51} (\bibinfo {year} {2021})},\ \Eprint
  {http://arxiv.org/abs/2002.12347} {arXiv:2002.12347 [hep-lat]} \BibitemShut
  {NoStop}%
\bibitem [{\citenamefont {C\`e}\ \emph {et~al.}(2022)\citenamefont {C\`e} \emph
  {et~al.}}]{Ce:2022kxy}%
  \BibitemOpen
  \bibfield  {author} {\bibinfo {author} {\bibfnamefont {M.}~\bibnamefont
  {C\`e}} \emph {et~al.},\ }\href {\doibase 10.1103/PhysRevD.106.114502}
  {\bibfield  {journal} {\bibinfo  {journal} {Phys. Rev. D}\ }\textbf {\bibinfo
  {volume} {106}},\ \bibinfo {pages} {114502} (\bibinfo {year} {2022})},\
  \Eprint {http://arxiv.org/abs/2206.06582} {arXiv:2206.06582 [hep-lat]}
  \BibitemShut {NoStop}%
\bibitem [{\citenamefont {Alexandrou}\ \emph {et~al.}(2022)\citenamefont
  {Alexandrou} \emph {et~al.}}]{Alexandrou:2022amy}%
  \BibitemOpen
  \bibfield  {author} {\bibinfo {author} {\bibfnamefont {C.}~\bibnamefont
  {Alexandrou}} \emph {et~al.},\ }\href@noop {} {\  (\bibinfo {year} {2022})},\
  \Eprint {http://arxiv.org/abs/2206.15084} {arXiv:2206.15084 [hep-lat]}
  \BibitemShut {NoStop}%
\bibitem [{\citenamefont {Colangelo}\ \emph {et~al.}(2022)\citenamefont
  {Colangelo}, \citenamefont {El-Khadra}, \citenamefont {Hoferichter},
  \citenamefont {Keshavarzi}, \citenamefont {Lehner}, \citenamefont {Stoffer},\
  and\ \citenamefont {Teubner}}]{Colangelo:2022vok}%
  \BibitemOpen
  \bibfield  {author} {\bibinfo {author} {\bibfnamefont {G.}~\bibnamefont
  {Colangelo}}, \bibinfo {author} {\bibfnamefont {A.~X.}\ \bibnamefont
  {El-Khadra}}, \bibinfo {author} {\bibfnamefont {M.}~\bibnamefont
  {Hoferichter}}, \bibinfo {author} {\bibfnamefont {A.}~\bibnamefont
  {Keshavarzi}}, \bibinfo {author} {\bibfnamefont {C.}~\bibnamefont {Lehner}},
  \bibinfo {author} {\bibfnamefont {P.}~\bibnamefont {Stoffer}}, \ and\
  \bibinfo {author} {\bibfnamefont {T.}~\bibnamefont {Teubner}},\ }\href
  {\doibase 10.1016/j.physletb.2022.137313} {\bibfield  {journal} {\bibinfo
  {journal} {Phys. Lett. B}\ }\textbf {\bibinfo {volume} {833}},\ \bibinfo
  {pages} {137313} (\bibinfo {year} {2022})},\ \Eprint
  {http://arxiv.org/abs/2205.12963} {arXiv:2205.12963 [hep-ph]} \BibitemShut
  {NoStop}%
\bibitem [{\citenamefont {Lehner}(2022)}]{talkLehner2022}%
  \BibitemOpen
  \bibfield  {author} {\bibinfo {author} {\bibfnamefont {C.}~\bibnamefont
  {Lehner}} (\bibinfo {collaboration} {RBC and UKQCD}),\ }\href@noop {}
  {\enquote {\bibinfo {title} {The hadronic vacuum polarization},}\ }\bibinfo
  {howpublished}
  {\url{https://indico.ph.ed.ac.uk/event/112/contributions/1660/attachments/1000/1391/talk-nobackup.pdf}}
  (\bibinfo {year} {2022}),\ \bibinfo {note} {fifth Plenary Workshop of the
  Muon g-2 Theory Initiative, Edinburgh, UK}\BibitemShut {NoStop}%
\bibitem [{\citenamefont {Gottlieb}(2022)}]{talkGottlieb2022}%
  \BibitemOpen
  \bibfield  {author} {\bibinfo {author} {\bibfnamefont {S.}~\bibnamefont
  {Gottlieb}} (\bibinfo {collaboration} {Fermilab Lattice, HPQCD and MILC}),\
  }\href@noop {} {\enquote {\bibinfo {title} {Hadronic vacuum polarization: An
  unblinded window on the g-2 mystery},}\ }\bibinfo {howpublished}
  {\url{https://www.benasque.org/2022lattice_workshop/talks_contr/158_Gottlieb_gm2_LatticeNET.pdf}}
  (\bibinfo {year} {2022}),\ \bibinfo {note} {first LatticeNET Workshop on
  challenges in Lattice field theory, Benasque, Spain}\BibitemShut {NoStop}%
\bibitem [{\citenamefont {Colangelo}(2022)}]{talkColangelo2022}%
  \BibitemOpen
  \bibfield  {author} {\bibinfo {author} {\bibfnamefont {G.}~\bibnamefont
  {Colangelo}},\ }\href@noop {} {\enquote {\bibinfo {title} {Dispersive
  calculation of hadronic contributions to $(g-2)_\mu$},}\ }\bibinfo
  {howpublished}
  {\url{https://www.benasque.org/2022lattice_workshop/talks_contr/153_g-2_Benasque-2022.pdf}}
  (\bibinfo {year} {2022}),\ \bibinfo {note} {first LatticeNET Workshop on
  challenges in Lattice field theory, Benasque, Spain}\BibitemShut {NoStop}%
\end{thebibliography}%

\end{document}